\documentclass[prl,twocolumn,superscriptaddress]{revtex4-1}

\AtBeginDocument{\renewcommand{\natexlab}[1]{#1}}

\usepackage[dvipdfmx]{graphicx}
\usepackage{dcolumn}
\usepackage{bm}
\usepackage{color}

\graphicspath{{fig/}}

\begin{document}

\preprint{APS/123-QED}

\title{
Model construction 
and a possibility of cuprate-like pairing in a new $d^9$ nickelate superconductor (Nd,Sr)NiO$_2$
} 
\author{Hirofumi Sakakibara}
\email{sakakibara.tottori.u@gmail.com}
\affiliation{Department of Applied Mathematics and Physics, Tottori University, Tottori, Tottori 680-8552, Japan}
\affiliation{Computational Condensed Matter Physics Laboratory, RIKEN, Wako, Saitama 351-0198, Japan}
\author{Hidetomo Usui}
\affiliation{Department of Physics and Materials Science, Shimane University, 1060 Nishikawatsu-cho, Matsue, Shimane 690-8504, Japan}
\author{Katsuhiro Suzuki}
\affiliation{Research Organization of Science and Technology, Ritsumeikan University, Kusatsu, Shiga 525-8577, Japan}
\author{Takao Kotani}
\affiliation{Department of Applied Mathematics and Physics, Tottori University, Tottori, Tottori 680-8552, Japan}
\author{Hideo Aoki}
\affiliation{National Institute of Advanced Industrial Science and Technology (AIST), Tsukuba, Ibaraki 305-8568, Japan}
\affiliation{Department of Physics, The University of Tokyo, Hongo, Tokyo 113-0033, Japan}
\author{Kazuhiko Kuroki}
\affiliation{Department of Physics, Osaka University, 1-1 Machikaneyama-cho, Toyonaka, Osaka 560-0043, Japan}

\date{\today}

\begin{abstract}
Effective models are constructed for a newly discovered superconductor (Nd,Sr)NiO$_2$, which has been considered as a possible nickelate analogue of the cuprates. Estimation of the effective interaction, which turns out to require a multiorbital model that takes account of all the orbitals involved on the Fermi surface, shows that the effective interactions are significantly larger than in the cuprates. A fluctuation exchange study suggests occurrence of $d_{x^2-y^2}$-wave superconductivity, where the transition temperature 
can be lowered from the cuprates due to the larger interaction.
\end{abstract}

\pacs{ }
\maketitle
{\it Introduction -} 
While it exceeds more than three decades since the high $T_c$ superconductivity was discovered in the cuprates, search for their analogues in non-copper-based materials has remained a big challenge, both experimentally and theoretically. In particular, nickelates have attracted attention due to their electronic configuration close to the cuprates. For instance, LaNiO$_3$/LaAlO$_3$ superlattice has been proposed as a possible candidate. There, Ni takes the 3+ valence with $d^7$ configuration, and the $d_{x^2-y^2}$ orbital is lowered in energy below $d_{3z^2-r^2}$, 
resulting in a single electron occupation of the $d_{x^2-y^2}$ orbital \cite{Chaloupka,Hansmann,Han}.  
Other materials considered as having electronic states close to the cuprates are multilayer nickelates $Ln_{n+1}$Ni$_n$O$_{2(n+1)}$ ($Ln=$La, Nd, Pr) without apical oxygens    
 \cite{JZhang,PardoPickett,JGCheng,Retoux,Warren,Poltavets438,Poltavets326,Bernal}, where the Ni $3d_{x^2-y^2}$ band is expected to approach half-filled as the number of layers $n$ increases.  In particular, the infinite-layer nickelates ($Ln$NiO$_2$) are of special interest because Ni$^{1+}$ valence, hence $d^9$ configuration, is expected if we assume $Ln^{3+}$ and O$^{2-}$ valence \cite{Crespin,Hayward,HaywardNd,Kawai,Kaneko,Ikeda,Onozuka}. First-principles studies on LaNiO$_2$ have pointed out similarities as well as differences from the cuprates \cite{Anisimov,Pickett}. 
In Ref. \onlinecite{Anisimov}, it was found that the layered structure without the apical oxygens can favor a low-spin state when holes are doped, as in the cuprates. Also, these first-principles 
studies predict antiferromagnetic ordering (but with small energy gain from the paramagnetic state \cite{Pickett}), while no magnetism is observed 
experimentally \cite{Hayward,HaywardNd}. Superconductivity, despite many years of challenge, had not been found till very recently, but superconductivity with $T_c=9\sim 15$ K has finally been discovered in a Nd$_{0.8}$Sr$_{0.2}$NiO$_2$ thin film synthesized on SrTiO$_3$ substrate \cite{Hwang,Sawatzky}. Now we have a theoretical 
challenge to grasp the material's electronic structure and to resolve some important puzzles (on the mother compound 
being metallic without magnetism, and $T_c$ lower than in the cuprates).

This precisely motivates the present study, where we first construct effective low-energy models for the infinite-layer nickelate 
to compare them with that for a high-$T_c$ cuprate superconductor HgBa$_2$CuO$_4$. For the mother (undoped) nickelate, we shall show that relatively small amount of holes are {\it self-doped} into the Ni $3d_{x^2-y^2}$ orbital due to the presence of La-originated electron pockets, which is 
likely to prevent the $3d_{x^2-y^2}$ band from being in a Mott insulating state.  When we turn to the Sr-doped case, we shall find that the occurrence of $d_{x^2-y^2}$-wave superconductivity is suggested as in the cuprates \cite{KeimerReview}, where a large intra-orbital interaction  (denoted as $U_{d_{x^2-y^2}}$) within the $d_{x^2-y^2}$ orbitals will suppress $T_c$ due to strong renormalization effects.

The model construction is done in three steps.
We start with a first-principles calculation with the local density approximation (LDA) using {\sf ecalj} package \cite{ecalj}. 
We then obtain the model parameters in the one-body Hamiltonian in terms of the standard 
maximally-localized Wannier functions \cite{Marzari,Souza}.
In the final step, we obtain the 
model parameters in the many-body Hamiltonian 
with the constrained random-phase approximation (cRPA) \cite{Aryasetiawan_2004},
where we use the tetrahedron method \cite{kotani_quasiparticle_2007,kotani_quasiparticle_2014} 
for Brillouin-zone sampling \cite{comment1}.

We then explore possibility of superconductivity and tendency toward magnetism 
for the obtained low-energy models with the fluctuation-exchange approximation (FLEX)\cite{Bickers1989,Bickers1991,Ikeda_omega0,comment7}, 
where we only consider the on-site interactions. 
The obtained Green's function and the pairing interaction, mediated mainly by spin fluctuations, are plugged into the linearized Eliashberg equation.  
We adopt the eigenvalue $\lambda$ of the Eliashberg equation as a measure of superconductivity, and 
the spin Stoner factor $\alpha_S$, 
given as the maximum eigenvalue of the 
product between the bare Coulomb interaction in the spin channel and the irreducible susceptibility
$\hat \chi_0$ (see, e.g., \onlinecite{Sakakibara2}), as 
a measure of antiferromagnetism 
(with $\lambda=1$ and $\alpha_S=1$ signalling superconductivity and magnetic ordering, respectively). Throughout the study, the eigenfunction of the Eliashberg equation that has the largest $\lambda$ always has $d_{x^2-y^2}$-wave pairing symmetry (see the inset of Figure \ref{fig3}(a)). 
We take 8$\times$8$\times$8 (32$\times$32$\times$2) $k$-point mesh and 8192 (4096) Matsubara frequencies for the FLEX calculation of the nickelate (cuprate).

\begin{figure}
	\includegraphics[width=9cm]{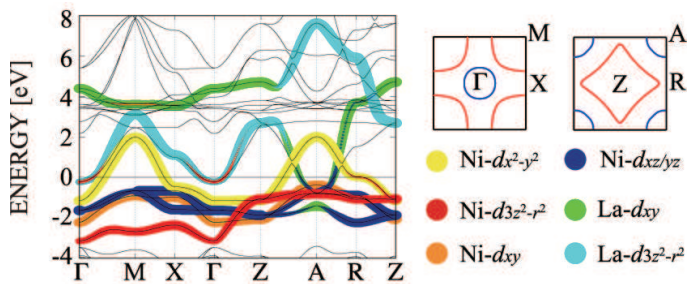}
	\caption{ First-principles band structure of LaNiO$_2$ (blue solid lines). The band structure of the seven-orbital model is superposed, where 
the Wannier-orbital weight is represented by the thickness of lines with color-coded orbital characters.
	Inset shows cross sections of the Fermi surface at $k_z=0$ (left) and $k_z=\pi$ (right), where the red and blue lines depict Ni- and La-originated ones, respectively.See the supplemental material for the 3D plot of the Fermi surface.}
	\label{fig1}
\end{figure}

{\it Mother nickelate -} 
We first perform first-principles calculation for the mother compound LaNiO$_2$ adopting the lattice parameters determined for NdNiO$_2$ in Ref. \onlinecite{HaywardNd}. Here we consider LaNiO$_2$ instead of NdNiO$_2$ itself to avoid ambiguity for the treatment of the $f$-orbital bands. We show in the supplemental material that LaNiO$_2$ and NdNiO$_2$ in fact 
give essentially the same band structure (except for the $f$ bands) if we adopt the same set of lattice parameters. The obtained band structure is displayed in Fig. \ref{fig1}, which is similar to that obtained for LaNiO$_2$ in previous studies \cite{Anisimov,Pickett}. 
A prominent feature, as compared to the cuprates, is that, on top of the main Ni $3d_{x^2-y^2}$ band, other bands that have La $5d$ character, mixed with Ni $3d$, intersect the Fermi level.  
This La-originated electron pockets may be an origin of the experimentally observed metallic behavior of the resistivity at high temperatures as well as the negative Hall coefficient\cite{Ikeda,Hwang}.  
The presence of the La-originated Fermi surface
also suggests that holes should be self-doped into the Ni $3d$ orbitals. 

We now construct a low-energy model from the first-principles bands around the Fermi level. Here, we aim to construct a model that explicitly considers the Ni- 
and  La-centered Wannier orbitals. Ni $3d_{x^2-y^2}$, Ni $3d_{3z^2-r^2}$, Ni $3d_{xz}$, Ni $3d_{yz}$, La $5d_{xy}$, and La $5d_{3z^2-r^2}$ are known to have weights on the Fermi surface\cite{Pickett}. In addition, here we opt to include the Ni $3d_{xy}$ orbital, whose band actually lies closer to the Fermi level than Ni $3d_{3z^2-r^2}$ in some portions of the Brillouin zone. In fact, we notice that the inclusion of the Ni $3d_{xy}$ is crucial for stabilizing the Wannierization procedure, although it does not contribute to the Fermi surface. Another possible way to construct a model is to explicitly consider the oxygen $2p$ orbitals. We shall actually construct such models for discussions on electronic structures, while for many-body calculations such models have too  many orbitals. So we mainly restrict ourselves to the above seven-orbital model, which still takes account of the oxygen orbitals through the Wannier orbitals implicitly. 

In Fig. \ref{fig1}, the band structure of the seven-orbital model is superposed to the first-principles band structure. 
The estimated values of the on-site interactions are listed in Table \ref{tab1}. 
We shall later compare these with those in the cuprates.

\begin{table}[!h]
\caption{The on-site interactions for the mother compound and $p=0.2$ doped one evaluated with cRPA.  $U$ ($U'$) are 
the intra-orbital (inter-orbital) Coulomb repulsions and $J$ the Hund's coupling. $U$ for the seven-orbital model is given for Ni $3d_{x^2-y^2}$, $3d_{3z^2-r^2}(\equiv 3d_{z^2})$ and La $5d_{xy}$, $5d_{3z^2-r^2}(\equiv 5d_{z^2})$ orbitals, and 
$U'$ and $J$ are those between these orbitals.
Interactions for HgBa$_2$CuO$_4$ estimated 
for the five-orbital model are also listed for comparison,
where $U'$ and $J$ are those between 6$s$ and 6$p$ orbitals. 
}
\label{tab1}
\begin{tabular}{ c c | c  | c c|| c} \hline
 &  & LaNiO$_2$    &    LaNiO$_2$  &    ($p$=0.2)   &HgBa$_2$CuO$_4$    \\
 &[eV]&  7orbital&  7orbital  & 2orbital  & 5orbital \\\hline
 & $U_{d_{x^2-y^2}}$ & 3.81 & 	4.19 & 2.57  &  2.60\\
Ni/Cu&$U_{d_{z^2}}$ & 4.55   &  5.26  & 2.57 &  5.96 \\
(3$d$)&$U'$	           & 2.62 &	3.13  & 1.25 & 2.50 \\
&$J$	                 & 0.71  &	0.73   &  0.52& 0.63 \\\hline 
&$U_{d_{xy}}/U_{s}$                & 1.99  &     2.25   &  --&   2.82\\\
La/Hg&$U_{d_{z^2}}/U_{p_{x,y}}$  &1.78    &	2.05   & --  &  2.22\\
(5$d$/$6s,6p$)&$U'$	             & 1.52   &	1.78   & --  & 1.87\\
&$J$	             & 0.37  &	0.38   & -- & 0.22\\
\hline
\end{tabular}
\end{table}

From charge neutrality, the total density of  electrons 
in the seven-orbital model is 9 electrons per unit cell. 
The orbital-resolved density is estimated to be $n_{{\rm Ni} d_{x^2-y^2}}=0.94$, $n_{{\rm Ni}  d_{3z^2-r^2}}=1.83$, $n_{{\rm Ni} d_{xy}}=1.97$, $n_{{\rm Ni} d_{xz}+d_{yz}}=3.89$, 
$n_{{\rm La} d_{3z^2-r^2}}=0.12$, and $n_{{\rm La} d_{xy}}=0.25$.
If it were not for the bands having the La character, the $d^9$ configuration would give $n_{{\rm Ni} d_{x^2-y^2}}=1.0$. The present result shows that about 0.06 holes per unit cell exist in the Ni $3d_{x^2-y^2}$ orbital that are self-doped from the La electron pockets.

It is thus likely that the self-doping prevents the Ni $3d_{x^2-y^2}$ band from being in a Mott insulating state.  In addition, the Fermi surface of the Ni $3d_{x^2-y^2}$ band is strongly warped and its nesting is not so good, so that the tendency toward magnetic ordering may not be strong. In fact, previous first-principles studies predict that, although antiferromagnetism exists in LaNiO$_2$ \cite{Anisimov,Pickett}, the energy gain from the paramagnetic state is small \cite{Pickett}. In the actual materials, magnetic long-range order is observed neither in LaNiO$_2$ nor NdNiO$_2$, which has been attributed to the Ni$^{2+}$ centers due to excess oxygens as well as structural disorder present in the actual materials \cite{Hayward,HaywardNd}.  The present model construction suggests that the absence of the Mott insulating state together with the bad nesting may result in the absence of magnetic ordering even in a ideal, stoichiometric material. This sharply contrasts with the cuprates, where the mother compounds are Mott insulators.  
The mother nickelate, despite its metallicity, 
is not superconducting, for which we speculate 
should be because the electronic state of the Ni $3d_{x^2-y^2}$ band, with only a small amount of doped holes, resembles that of the heavily underdoped cuprates. The FLEX approximation cannot treat electron correlation effects in such 
a regime, so we will not analyze superconductivity here, 
and leave confirmation of this picture for future studies.

\begin{figure}
	\includegraphics[width=9cm]{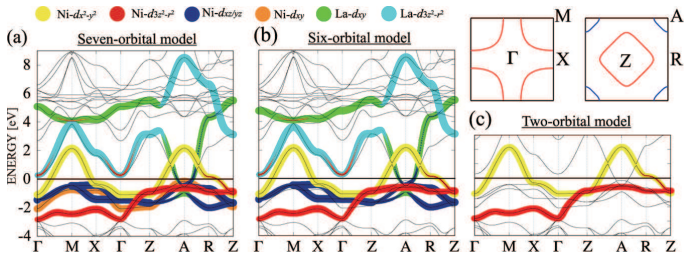}
	\caption{The band structure of the doped 
nickelate with  $p=0.2$  in (a) the seven-orbital model, (b) the six-orbital model (see text), and (c) the two-orbital model with color-coded orbital 
characters, superposed with the first-principles band structure (black lines). 
The cross sections of the Fermi surfaces are depicted in the inset as in Fig. \ref{fig1}.
}
	\label{fig2}
\end{figure}

\begin{figure}
	\includegraphics[width=8cm]{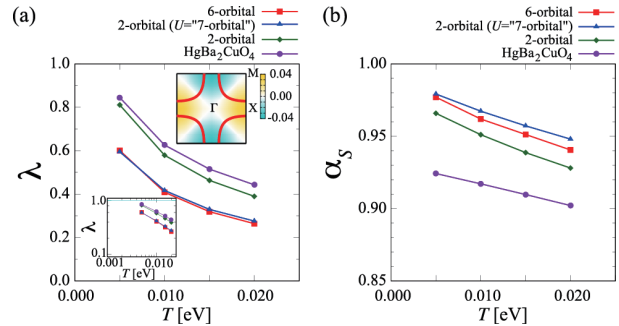}
	\caption{Temperature dependence of the $d_{x^2-y^2}$-wave eigenvalue $\lambda$ of the Eliashberg equation (a) and the Stoner factor $\alpha _S$ (b) for the six- and two-orbital models for $p=0.2$. Also shown is the result for the two-orbital model with the interaction parameters taken to be as in the seven-orbital model (indicated as $U$=``7-orbital'').  For comparison, result for 
HgBa$_2$CuO$_4$ in 
the five-orbital model (with the same $n_{3d_{x^2-y^2}}$ as in the $p=0.2$ nickelate) is also shown.  
The insets in (a) are a log-log plot of $\lambda$ vs. $T$, and the eigenfunction of the Eliashberg equation at $k_z=0$ (see the supplemental material for other $k_z$ cuts) for the six-orbital model of the nickelate ($p$=0.2) at $T=0.005$ eV.}
	\label{fig3}
\end{figure}

{\it Doped nickelate -} 
We next turn to the Sr-doped case, where superconductivity is observed experimentally \cite{Hwang}. Since the electron pockets in the mother compound has large La components, the rigid-band picture should be invalidated. Here we obtain the band structure using the virtual-crystal approximation (VCA), adopting the experimental lattice parameters of NdNiO$_2$\cite{HaywardNd}. Due to technical reasons in the VCA, we use Ba instead of Sr \cite{comment}. The calculation is performed for La$_{0.8}$Ba$_{0.2}$NiO$_2$ (denoted as $p=0.2$ hereafter), and we construct a seven-orbital model as in the mother compound. The first-principles band structure is seen to accurately agree with that of the seven-orbital model as displayed in Fig. \ref{fig2}(a).   The estimated interactions  are listed in Table \ref{tab1}.

Performing FLEX calculation for the seven-orbital model on a three-dimensional $k$-mesh would be tedious, especially at low temperatures. Since the Ni $3d_{xy}$ band barely hybridizes with the other bands, ignoring this orbital 
in the seven-orbital model hardly affects the band structure for the remaining six orbitals, as shown in Fig. \ref{fig2}(b).  Also, the Ni $d_{xy}$ band is fully-filled, so that we expect that removing this band does not affect the FLEX results. We have checked this by comparing the FLEX results for the seven and six-orbital models at $T=0.03$ eV, where we find basically the same results with 
$\lambda=0.214$, $\alpha_S=0.926$ for the seven-orbital, 
$\lambda=0.211$, $\alpha_S=0.926$ for the six-orbital \cite{comment6}.  This enables us to perform the FLEX calculation down to lower temperatures for the Ni-$d_{xy}$-eliminated six-orbital model, where we adopt the interaction parameters estimated for the seven-orbital model.

From the studies on the cuprates, the main player in the superconductivity in the nickelate is expected 
to be the Ni $3d_{x^2-y^2}$ band, which produces the main Fermi surface. To see if the orbitals that have no weight on the main Fermi surface have any effects on  superconductivity or the magnetism (apart from the self-doping effect), we further construct a two-orbital model (Fig. \ref{fig2}(c)), where only the Ni $3d_{x^2-y^2}$ and $3d_{3z^2-r^2}$ orbitals are explicitly taken into account \cite{Sakakibara1}. 
We can notice that the interaction parameters estimated for the two-orbital model, included in Table \ref{tab1}, are 
significantly reduced from the seven-orbital counterparts. This is because the La bands are metallic, so that their screening effect, when taken into account effectively in the two-orbital model, is substantial.

To make a quantitative comparison with the cuprates, we 
also construct a model for 
HgBa$_2$CuO$_4$ adopting the crystal structure determined in Ref. \onlinecite{Wagner}. The on-site interactions for the cuprates have been estimated in Refs. \onlinecite{Jang_crpa,mrpa,mrpa2} within the $d_{x^2-y^2}$-$d_{3z^2-r^2}$ two-orbital models, but here we construct, to make a fair comparison with the nickelate, a five-orbital model, where we explicitly take account also of the Hg $6s$, $6p_x$, and $6p_y$ orbitals, whose bands overlap with the Cu $3d_{x^2-y^2}$ band in energy(see the supplemental material). Although this effect is not as large as that for the La $5d$ orbitals in the nickelates, it still enhances the interactions appreciably, as seen by comparing the values given in Table \ref{tab1} with those in Ref. \onlinecite{Jang_crpa}. We have also checked that explicitly considering the $t_{2g}$ orbitals, which do not contribute to the Fermi surface in the cuprate, has small effect on the interaction values.
Then we can compare the nickelate and the cuprate, to realize that the interactions are significantly larger in the former, which should come from the smaller hybridization between the Ni and the oxygen atomic orbitals\cite{Anisimov,Pickett}. We will come back to this point later.

We now come to superconductivity. Figure \ref{fig3} 
compares the FLEX results for $\lambda$ ($d_{x^2-y^2}$-wave superconductivity) and $\alpha_S$ (magnetism) in the six-orbital 
 and two-orbital models for $p=0.2$, plotted against temperature. Since the electron pockets originating from the La orbitals are absent in the two-orbital model, we take account of the effect of self-doping there by setting the total density of electrons in such a way that $n_{{\rm Ni} 3d_{x^2-y^2}}$ equals the value determined from the seven-orbital model. We can immediately see that $\lambda$ is significantly reduced in the six-orbital model. This reduction of $\lambda$ in the six-orbital model can come from either larger values of the interaction, or the presence of the electron pockets. To identify which is the cause, we have performed another FLEX calculation for the two-orbital model adopting the same interaction values as in the seven-orbital model, as included in Fig. \ref{fig3}. The two-orbital model then gives results similar to those in the six-orbital model, which implies that the main origin of the reduction of $\lambda$ in the six-orbital model is the large renormalization effect due to the large $U_{d_{x^2-y^2}}$, rather than the presence of the electron pockets.  
So we can make an observation that the two-orbital model, even when the La orbitals are implicitly taken into 
account through the interaction parameters, 
is significantly 
inaccurate as far as the FLEX analysis is concerned 
(i.e., the screened interaction would result in an overestimated $\lambda$).  If we turn to 
the FLEX result for HgBa$_2$CuO$_4$ in the five-orbital model as included in Fig. \ref{fig3} for comparison, the cuprate exhibits larger  $\lambda$ than the nickelate, 
with smaller Stoner factor. This can again be attributed to the smaller $U_{d_{x^2-y^2}}$  in the cuprate.  
Hence the message here is that it is important to 
consider the electronic structure peculiar to the 
nickelate, where we have a larger $U_{d_{x^2-y^2}}$ along with a smaller bandwidth than in the cuprates, 
which is in turn responsible for the reduced $T_c$ 
through a strong renormalization effect. 

The larger interaction and the smaller bandwidth 
can be traced back to a larger level offset $\Delta_{dp}$ between $3d_{x^2-y^2}$ and oxygen $2p_{x,y}$ (atomic) orbitals for the nickelates than in the cuprates\cite{Anisimov,Pickett}. To evaluate $\Delta_{dp}$ quantitatively, we have also constructed models that explicitly consider the oxygen orbitals: a 23-orbital model (five Ni-$3d$, six O-$2p$, five La-$5d$, seven La-$4f$ orbitals) for LaNiO$_2$, and a 20-orbital model (five Cu-$3d$, twelve O-$2p$, two Hg-$6p$, and one Hg-$6s$ orbitals) for HgBa$_2$CuO$_4$, which gives $\Delta_{dp}=3.7$eV for the nickelate vs. $\Delta_{dp}=1.8$eV for the cuprate (see also a recent paper \cite{Botana}). The suppression of superconductivity due to large $\Delta_{dp}$ has been pointed out in previous studies for the cuprates\cite{Shinkai,Weber,Weber2}. The reduction of $T_c$ for large $\Delta_{dp}$ may also be viewed in terms of the strong-coupling picture, where the superexchange interaction is given as $J\propto t_{dp}^4/\Delta_{dp}^3$ with $t_{dp}$ being the nearest-neighbor $d_{x^2-y^2}$-$p_{x,y}$ hopping. If we go back to the oxygen-$2p$-integrated-out model taking the strong-coupling viewpoint, the nearest-neighbor spin-spin interaction is estimated as $J\propto t^2/U_{d_{x^2-y^2}}$, where $t$ is the nearest-neighbor hopping between $d_{x^2-y^2}$ orbitals. The $T_c$ reduction for large $U_{d_{x^2-y^2}}$ and small $t$ may also viewed in this way.

Conversely, the present study allows us to expect that 
reduction of the in-plane lattice constant, which will 
increase the Ni $d_{x^2-y^2}$ bandwidth and reduce the interaction within the Ni $d_{x^2-y^2}$ orbital, should enhance superconductivity.  Hence the smaller lattice constant in NdNiO$_2$ than in LaNiO$_2$ \cite{Hayward,HaywardNd} may be 
relevant in the observation of superconductivity only in the former so far. Applying physical pressure can thus be a route toward observation of positive effects on superconductivity in the nickelate family. 

{\it Summary -} We have constructed effective models for the newly discovered nickelate superconductor. For the mother compound, small amount of holes are self-doped into the Ni $3d_{x^2-y^2}$ orbital due to the presence of electron pockets. These electron pockets may have relevance to the metallic behavior observed experimentally \cite{Ikeda,Hwang}, while the electronic state of the hole self-doped $3d_{x^2-y^2}$ band may be close to that of the heavily underdoped cuprates \cite{KeimerReview}, without magnetism or superconductivity. For the Sr doped nickelate, the FLEX study for the six-orbital model that incorporates La orbitals as well indicates that $d_{x^2-y^2}$-wave superconductivity is suggested to arise as in the cuprates, but that a larger interaction within the Ni $d_{x^2-y^2}$ orbital along with a narrow bandwidth than in the cuprates results in a lower transition temperature due to the strong self-energy renormalization effect. Important future problems include exploration of related materials with different elements and/or compositions, which may result in a possible enhancement of $T_c$.

\begin{acknowledgments}
KK and HA are supported by JSPS KAKENHI Grant Number JP18H01860.
The computing resource is supported by Computing System for Research in Kyushu University (ITO system), 
the supercomputer system in RIKEN (HOKUSAI), and the supercomputer system in ISSP (sekirei).
\end{acknowledgments}

\bibliography{lanio2}

\begin{thebibliography}{46}%
\makeatletter
\providecommand \@ifxundefined [1]{%
 \@ifx{#1\undefined}
}%
\providecommand \@ifnum [1]{%
 \ifnum #1\expandafter \@firstoftwo
 \else \expandafter \@secondoftwo
 \fi
}%
\providecommand \@ifx [1]{%
 \ifx #1\expandafter \@firstoftwo
 \else \expandafter \@secondoftwo
 \fi
}%
\providecommand \natexlab [1]{#1}%
\providecommand \enquote  [1]{``#1''}%
\providecommand \bibnamefont  [1]{#1}%
\providecommand \bibfnamefont [1]{#1}%
\providecommand \citenamefont [1]{#1}%
\providecommand \href@noop [0]{\@secondoftwo}%
\providecommand \href [0]{\begingroup \@sanitize@url \@href}%
\providecommand \@href[1]{\@@startlink{#1}\@@href}%
\providecommand \@@href[1]{\endgroup#1\@@endlink}%
\providecommand \@sanitize@url [0]{\catcode `\\12\catcode `\$12\catcode
  `\&12\catcode `\#12\catcode `\^12\catcode `\_12\catcode `\%12\relax}%
\providecommand \@@startlink[1]{}%
\providecommand \@@endlink[0]{}%
\providecommand \url  [0]{\begingroup\@sanitize@url \@url }%
\providecommand \@url [1]{\endgroup\@href {#1}{\urlprefix }}%
\providecommand \urlprefix  [0]{URL }%
\providecommand \Eprint [0]{\href }%
\providecommand \doibase [0]{http://dx.doi.org/}%
\providecommand \selectlanguage [0]{\@gobble}%
\providecommand \bibinfo  [0]{\@secondoftwo}%
\providecommand \bibfield  [0]{\@secondoftwo}%
\providecommand \translation [1]{[#1]}%
\providecommand \BibitemOpen [0]{}%
\providecommand \bibitemStop [0]{}%
\providecommand \bibitemNoStop [0]{.\EOS\space}%
\providecommand \EOS [0]{\spacefactor3000\relax}%
\providecommand \BibitemShut  [1]{\csname bibitem#1\endcsname}%
\let\auto@bib@innerbib\@empty
\bibitem [{\citenamefont {Chaloupka}\ and\ \citenamefont
  {Khaliullin}(2008)}]{Chaloupka}%
  \BibitemOpen
  \bibfield  {author} {\bibinfo {author} {\bibfnamefont {J.}~\bibnamefont
  {Chaloupka}}\ and\ \bibinfo {author} {\bibfnamefont {G.}~\bibnamefont
  {Khaliullin}},\ }\href {\doibase 10.1103/PhysRevLett.100.016404} {\bibfield
  {journal} {\bibinfo  {journal} {Phys. Rev. Lett.}\ }\textbf {\bibinfo
  {volume} {100}},\ \bibinfo {pages} {016404} (\bibinfo {year}
  {2008})}\BibitemShut {NoStop}%
\bibitem [{\citenamefont {Hansmann}\ \emph {et~al.}(2009)\citenamefont
  {Hansmann}, \citenamefont {Yang}, \citenamefont {Toschi}, \citenamefont
  {Khaliullin}, \citenamefont {Andersen},\ and\ \citenamefont
  {Held}}]{Hansmann}%
  \BibitemOpen
  \bibfield  {author} {\bibinfo {author} {\bibfnamefont {P.}~\bibnamefont
  {Hansmann}}, \bibinfo {author} {\bibfnamefont {X.}~\bibnamefont {Yang}},
  \bibinfo {author} {\bibfnamefont {A.}~\bibnamefont {Toschi}}, \bibinfo
  {author} {\bibfnamefont {G.}~\bibnamefont {Khaliullin}}, \bibinfo {author}
  {\bibfnamefont {O.~K.}\ \bibnamefont {Andersen}}, \ and\ \bibinfo {author}
  {\bibfnamefont {K.}~\bibnamefont {Held}},\ }\href {\doibase
  10.1103/PhysRevLett.103.016401} {\bibfield  {journal} {\bibinfo  {journal}
  {Phys. Rev. Lett.}\ }\textbf {\bibinfo {volume} {103}},\ \bibinfo {pages}
  {016401} (\bibinfo {year} {2009})}\BibitemShut {NoStop}%
\bibitem [{\citenamefont {Han}\ \emph {et~al.}(2011)\citenamefont {Han},
  \citenamefont {Wang}, \citenamefont {Marianetti},\ and\ \citenamefont
  {Millis}}]{Han}%
  \BibitemOpen
  \bibfield  {author} {\bibinfo {author} {\bibfnamefont {M.~J.}\ \bibnamefont
  {Han}}, \bibinfo {author} {\bibfnamefont {X.}~\bibnamefont {Wang}}, \bibinfo
  {author} {\bibfnamefont {C.~A.}\ \bibnamefont {Marianetti}}, \ and\ \bibinfo
  {author} {\bibfnamefont {A.~J.}\ \bibnamefont {Millis}},\ }\href {\doibase
  10.1103/PhysRevLett.107.206804} {\bibfield  {journal} {\bibinfo  {journal}
  {Phys. Rev. Lett.}\ }\textbf {\bibinfo {volume} {107}},\ \bibinfo {pages}
  {206804} (\bibinfo {year} {2011})}\BibitemShut {NoStop}%
\bibitem [{\citenamefont {Zhang}\ \emph {et~al.}(2017)\citenamefont {Zhang},
  \citenamefont {Botana}, \citenamefont {Freeland}, \citenamefont {Phelan},
  \citenamefont {Zheng}, \citenamefont {Pardo}, \citenamefont {Norman},\ and\
  \citenamefont {Mitchell}}]{JZhang}%
  \BibitemOpen
  \bibfield  {author} {\bibinfo {author} {\bibfnamefont {J.}~\bibnamefont
  {Zhang}}, \bibinfo {author} {\bibfnamefont {A.~S.}\ \bibnamefont {Botana}},
  \bibinfo {author} {\bibfnamefont {J.~W.}\ \bibnamefont {Freeland}}, \bibinfo
  {author} {\bibfnamefont {D.}~\bibnamefont {Phelan}}, \bibinfo {author}
  {\bibfnamefont {H.}~\bibnamefont {Zheng}}, \bibinfo {author} {\bibfnamefont
  {V.}~\bibnamefont {Pardo}}, \bibinfo {author} {\bibfnamefont {M.~R.}\
  \bibnamefont {Norman}}, \ and\ \bibinfo {author} {\bibfnamefont {J.~F.}\
  \bibnamefont {Mitchell}},\ }\href {https://doi.org/10.1038/nphys4149
  10.1038/nphys4149
  https://www.nature.com/articles/nphys4149{\#}supplementary-information}
  {\bibfield  {journal} {\bibinfo  {journal} {Nature Physics}\ }\textbf
  {\bibinfo {volume} {13}},\ \bibinfo {pages} {864} (\bibinfo {year}
  {2017})}\BibitemShut {NoStop}%
\bibitem [{\citenamefont {Pardo}\ and\ \citenamefont
  {Pickett}(2010)}]{PardoPickett}%
  \BibitemOpen
  \bibfield  {author} {\bibinfo {author} {\bibfnamefont {V.}~\bibnamefont
  {Pardo}}\ and\ \bibinfo {author} {\bibfnamefont {W.~E.}\ \bibnamefont
  {Pickett}},\ }\href {\doibase 10.1103/PhysRevLett.105.266402} {\bibfield
  {journal} {\bibinfo  {journal} {Phys. Rev. Lett.}\ }\textbf {\bibinfo
  {volume} {105}},\ \bibinfo {pages} {266402} (\bibinfo {year}
  {2010})}\BibitemShut {NoStop}%
\bibitem [{\citenamefont {Cheng}\ \emph {et~al.}(2012)\citenamefont {Cheng},
  \citenamefont {Zhou}, \citenamefont {Goodenough}, \citenamefont {Zhou},
  \citenamefont {Matsubayashi}, \citenamefont {Uwatoko}, \citenamefont {Kong},
  \citenamefont {Jin}, \citenamefont {Yang},\ and\ \citenamefont
  {Shen}}]{JGCheng}%
  \BibitemOpen
  \bibfield  {author} {\bibinfo {author} {\bibfnamefont {J.-G.}\ \bibnamefont
  {Cheng}}, \bibinfo {author} {\bibfnamefont {J.-S.}\ \bibnamefont {Zhou}},
  \bibinfo {author} {\bibfnamefont {J.~B.}\ \bibnamefont {Goodenough}},
  \bibinfo {author} {\bibfnamefont {H.~D.}\ \bibnamefont {Zhou}}, \bibinfo
  {author} {\bibfnamefont {K.}~\bibnamefont {Matsubayashi}}, \bibinfo {author}
  {\bibfnamefont {Y.}~\bibnamefont {Uwatoko}}, \bibinfo {author} {\bibfnamefont
  {P.~P.}\ \bibnamefont {Kong}}, \bibinfo {author} {\bibfnamefont {C.~Q.}\
  \bibnamefont {Jin}}, \bibinfo {author} {\bibfnamefont {W.~G.}\ \bibnamefont
  {Yang}}, \ and\ \bibinfo {author} {\bibfnamefont {G.~Y.}\ \bibnamefont
  {Shen}},\ }\href {\doibase 10.1103/PhysRevLett.108.236403} {\bibfield
  {journal} {\bibinfo  {journal} {Phys. Rev. Lett.}\ }\textbf {\bibinfo
  {volume} {108}},\ \bibinfo {pages} {236403} (\bibinfo {year}
  {2012})}\BibitemShut {NoStop}%
\bibitem [{\citenamefont {Retoux}\ \emph {et~al.}(1998)\citenamefont {Retoux},
  \citenamefont {Rodriguez-Carvajal},\ and\ \citenamefont {Lacorre}}]{Retoux}%
  \BibitemOpen
  \bibfield  {author} {\bibinfo {author} {\bibfnamefont {R.}~\bibnamefont
  {Retoux}}, \bibinfo {author} {\bibfnamefont {J.}~\bibnamefont
  {Rodriguez-Carvajal}}, \ and\ \bibinfo {author} {\bibfnamefont
  {P.}~\bibnamefont {Lacorre}},\ }\href {\doibase
  https://doi.org/10.1006/jssc.1998.7892} {\bibfield  {journal} {\bibinfo
  {journal} {Journal of Solid State Chemistry}\ }\textbf {\bibinfo {volume}
  {140}},\ \bibinfo {pages} {307 } (\bibinfo {year} {1998})}\BibitemShut
  {NoStop}%
\bibitem [{\citenamefont {ApRoberts-Warren}\ \emph {et~al.}(2011)\citenamefont
  {ApRoberts-Warren}, \citenamefont {Dioguardi}, \citenamefont {Poltavets},
  \citenamefont {Greenblatt}, \citenamefont {Klavins},\ and\ \citenamefont
  {Curro}}]{Warren}%
  \BibitemOpen
  \bibfield  {author} {\bibinfo {author} {\bibfnamefont {N.}~\bibnamefont
  {ApRoberts-Warren}}, \bibinfo {author} {\bibfnamefont {A.~P.}\ \bibnamefont
  {Dioguardi}}, \bibinfo {author} {\bibfnamefont {V.~V.}\ \bibnamefont
  {Poltavets}}, \bibinfo {author} {\bibfnamefont {M.}~\bibnamefont
  {Greenblatt}}, \bibinfo {author} {\bibfnamefont {P.}~\bibnamefont {Klavins}},
  \ and\ \bibinfo {author} {\bibfnamefont {N.~J.}\ \bibnamefont {Curro}},\
  }\href {\doibase 10.1103/PhysRevB.83.014402} {\bibfield  {journal} {\bibinfo
  {journal} {Phys. Rev. B}\ }\textbf {\bibinfo {volume} {83}},\ \bibinfo
  {pages} {014402} (\bibinfo {year} {2011})}\BibitemShut {NoStop}%
\bibitem [{\citenamefont {Poltavets}\ \emph {et~al.}(2010)\citenamefont
  {Poltavets}, \citenamefont {Lokshin}, \citenamefont {Nevidomskyy},
  \citenamefont {Croft}, \citenamefont {Tyson}, \citenamefont {Hadermann},
  \citenamefont {Van~Tendeloo}, \citenamefont {Egami}, \citenamefont {Kotliar},
  \citenamefont {ApRoberts-Warren}, \citenamefont {Dioguardi}, \citenamefont
  {Curro},\ and\ \citenamefont {Greenblatt}}]{Poltavets438}%
  \BibitemOpen
  \bibfield  {author} {\bibinfo {author} {\bibfnamefont {V.~V.}\ \bibnamefont
  {Poltavets}}, \bibinfo {author} {\bibfnamefont {K.~A.}\ \bibnamefont
  {Lokshin}}, \bibinfo {author} {\bibfnamefont {A.~H.}\ \bibnamefont
  {Nevidomskyy}}, \bibinfo {author} {\bibfnamefont {M.}~\bibnamefont {Croft}},
  \bibinfo {author} {\bibfnamefont {T.~A.}\ \bibnamefont {Tyson}}, \bibinfo
  {author} {\bibfnamefont {J.}~\bibnamefont {Hadermann}}, \bibinfo {author}
  {\bibfnamefont {G.}~\bibnamefont {Van~Tendeloo}}, \bibinfo {author}
  {\bibfnamefont {T.}~\bibnamefont {Egami}}, \bibinfo {author} {\bibfnamefont
  {G.}~\bibnamefont {Kotliar}}, \bibinfo {author} {\bibfnamefont
  {N.}~\bibnamefont {ApRoberts-Warren}}, \bibinfo {author} {\bibfnamefont
  {A.~P.}\ \bibnamefont {Dioguardi}}, \bibinfo {author} {\bibfnamefont {N.~J.}\
  \bibnamefont {Curro}}, \ and\ \bibinfo {author} {\bibfnamefont
  {M.}~\bibnamefont {Greenblatt}},\ }\href {\doibase
  10.1103/PhysRevLett.104.206403} {\bibfield  {journal} {\bibinfo  {journal}
  {Phys. Rev. Lett.}\ }\textbf {\bibinfo {volume} {104}},\ \bibinfo {pages}
  {206403} (\bibinfo {year} {2010})}\BibitemShut {NoStop}%
\bibitem [{\citenamefont {Poltavets}\ \emph {et~al.}(2006)\citenamefont
  {Poltavets}, \citenamefont {Lokshin}, \citenamefont {Dikmen}, \citenamefont
  {Croft}, \citenamefont {Egami},\ and\ \citenamefont
  {Greenblatt}}]{Poltavets326}%
  \BibitemOpen
  \bibfield  {author} {\bibinfo {author} {\bibfnamefont {V.~V.}\ \bibnamefont
  {Poltavets}}, \bibinfo {author} {\bibfnamefont {K.~A.}\ \bibnamefont
  {Lokshin}}, \bibinfo {author} {\bibfnamefont {S.}~\bibnamefont {Dikmen}},
  \bibinfo {author} {\bibfnamefont {M.}~\bibnamefont {Croft}}, \bibinfo
  {author} {\bibfnamefont {T.}~\bibnamefont {Egami}}, \ and\ \bibinfo {author}
  {\bibfnamefont {M.}~\bibnamefont {Greenblatt}},\ }\href {\doibase
  10.1021/ja063031o} {\bibfield  {journal} {\bibinfo  {journal} {Journal of the
  American Chemical Society}\ }\textbf {\bibinfo {volume} {128}},\ \bibinfo
  {pages} {9050} (\bibinfo {year} {2006})},\ \bibinfo {note} {pMID: 16834375},\
  \Eprint {http://arxiv.org/abs/https://doi.org/10.1021/ja063031o}
  {https://doi.org/10.1021/ja063031o} \BibitemShut {NoStop}%
\bibitem [{\citenamefont {Bernal}\ \emph {et~al.}(2019)\citenamefont {Bernal},
  \citenamefont {MacLaughlin}, \citenamefont {Morris}, \citenamefont {Ho},
  \citenamefont {Shu}, \citenamefont {Tan}, \citenamefont {Zhang},
  \citenamefont {Ding}, \citenamefont {Huang},\ and\ \citenamefont
  {Poltavets}}]{Bernal}%
  \BibitemOpen
  \bibfield  {author} {\bibinfo {author} {\bibfnamefont {O.~O.}\ \bibnamefont
  {Bernal}}, \bibinfo {author} {\bibfnamefont {D.~E.}\ \bibnamefont
  {MacLaughlin}}, \bibinfo {author} {\bibfnamefont {G.~D.}\ \bibnamefont
  {Morris}}, \bibinfo {author} {\bibfnamefont {P.-C.}\ \bibnamefont {Ho}},
  \bibinfo {author} {\bibfnamefont {L.}~\bibnamefont {Shu}}, \bibinfo {author}
  {\bibfnamefont {C.}~\bibnamefont {Tan}}, \bibinfo {author} {\bibfnamefont
  {J.}~\bibnamefont {Zhang}}, \bibinfo {author} {\bibfnamefont
  {Z.}~\bibnamefont {Ding}}, \bibinfo {author} {\bibfnamefont {K.}~\bibnamefont
  {Huang}}, \ and\ \bibinfo {author} {\bibfnamefont {V.~V.}\ \bibnamefont
  {Poltavets}},\ }\href@noop {} {} (\bibinfo {year} {2019}),\ \Eprint
  {http://arxiv.org/abs/arXiv:1903.11706} {arXiv:1903.11706} \BibitemShut
  {NoStop}%
\bibitem [{\citenamefont {Crespin}\ \emph {et~al.}(1983)\citenamefont
  {Crespin}, \citenamefont {Levitz},\ and\ \citenamefont {Gatineau}}]{Crespin}%
  \BibitemOpen
  \bibfield  {author} {\bibinfo {author} {\bibfnamefont {M.}~\bibnamefont
  {Crespin}}, \bibinfo {author} {\bibfnamefont {P.}~\bibnamefont {Levitz}}, \
  and\ \bibinfo {author} {\bibfnamefont {L.}~\bibnamefont {Gatineau}},\ }\href
  {\doibase 10.1039/F29837901181} {\bibfield  {journal} {\bibinfo  {journal}
  {J. Chem. Soc.{,} Faraday Trans. 2}\ }\textbf {\bibinfo {volume} {79}},\
  \bibinfo {pages} {1181} (\bibinfo {year} {1983})}\BibitemShut {NoStop}%
\bibitem [{\citenamefont {Hayward}\ \emph {et~al.}(1999)\citenamefont
  {Hayward}, \citenamefont {Green}, \citenamefont {Rosseinsky},\ and\
  \citenamefont {Sloan}}]{Hayward}%
  \BibitemOpen
  \bibfield  {author} {\bibinfo {author} {\bibfnamefont {M.~A.}\ \bibnamefont
  {Hayward}}, \bibinfo {author} {\bibfnamefont {M.~A.}\ \bibnamefont {Green}},
  \bibinfo {author} {\bibfnamefont {M.~J.}\ \bibnamefont {Rosseinsky}}, \ and\
  \bibinfo {author} {\bibfnamefont {J.}~\bibnamefont {Sloan}},\ }\href
  {\doibase 10.1021/ja991573i} {\bibfield  {journal} {\bibinfo  {journal}
  {Journal of the American Chemical Society}\ }\textbf {\bibinfo {volume}
  {121}},\ \bibinfo {pages} {8843} (\bibinfo {year} {1999})},\ \Eprint
  {http://arxiv.org/abs/https://doi.org/10.1021/ja991573i}
  {https://doi.org/10.1021/ja991573i} \BibitemShut {NoStop}%
\bibitem [{\citenamefont {Hayward}\ and\ \citenamefont
  {Rosseinsky}(2003)}]{HaywardNd}%
  \BibitemOpen
  \bibfield  {author} {\bibinfo {author} {\bibfnamefont {M.}~\bibnamefont
  {Hayward}}\ and\ \bibinfo {author} {\bibfnamefont {M.}~\bibnamefont
  {Rosseinsky}},\ }\href {\doibase
  https://doi.org/10.1016/S1293-2558(03)00111-0} {\bibfield  {journal}
  {\bibinfo  {journal} {Solid State Sciences}\ }\textbf {\bibinfo {volume}
  {5}},\ \bibinfo {pages} {839 } (\bibinfo {year} {2003})},\ \bibinfo {note}
  {international Conference on Inorganic Materials 2002}\BibitemShut {NoStop}%
\bibitem [{\citenamefont {Kawai}\ \emph {et~al.}(2009)\citenamefont {Kawai},
  \citenamefont {Inoue}, \citenamefont {Mizumaki}, \citenamefont {Kawamura},
  \citenamefont {Ichikawa},\ and\ \citenamefont {Shimakawa}}]{Kawai}%
  \BibitemOpen
  \bibfield  {author} {\bibinfo {author} {\bibfnamefont {M.}~\bibnamefont
  {Kawai}}, \bibinfo {author} {\bibfnamefont {S.}~\bibnamefont {Inoue}},
  \bibinfo {author} {\bibfnamefont {M.}~\bibnamefont {Mizumaki}}, \bibinfo
  {author} {\bibfnamefont {N.}~\bibnamefont {Kawamura}}, \bibinfo {author}
  {\bibfnamefont {N.}~\bibnamefont {Ichikawa}}, \ and\ \bibinfo {author}
  {\bibfnamefont {Y.}~\bibnamefont {Shimakawa}},\ }\href {\doibase
  10.1063/1.3078276} {\bibfield  {journal} {\bibinfo  {journal} {Applied
  Physics Letters}\ }\textbf {\bibinfo {volume} {94}},\ \bibinfo {pages}
  {082102} (\bibinfo {year} {2009})},\ \Eprint
  {http://arxiv.org/abs/https://doi.org/10.1063/1.3078276}
  {https://doi.org/10.1063/1.3078276} \BibitemShut {NoStop}%
\bibitem [{\citenamefont {Kaneko}\ \emph {et~al.}(2009)\citenamefont {Kaneko},
  \citenamefont {Yamagishi}, \citenamefont {Tsukada}, \citenamefont {Manabe},\
  and\ \citenamefont {Naito}}]{Kaneko}%
  \BibitemOpen
  \bibfield  {author} {\bibinfo {author} {\bibfnamefont {D.}~\bibnamefont
  {Kaneko}}, \bibinfo {author} {\bibfnamefont {K.}~\bibnamefont {Yamagishi}},
  \bibinfo {author} {\bibfnamefont {A.}~\bibnamefont {Tsukada}}, \bibinfo
  {author} {\bibfnamefont {T.}~\bibnamefont {Manabe}}, \ and\ \bibinfo {author}
  {\bibfnamefont {M.}~\bibnamefont {Naito}},\ }\href {\doibase
  https://doi.org/10.1016/j.physc.2009.05.104} {\bibfield  {journal} {\bibinfo
  {journal} {Physica C: Superconductivity}\ }\textbf {\bibinfo {volume}
  {469}},\ \bibinfo {pages} {936 } (\bibinfo {year} {2009})},\ \bibinfo {note}
  {proceedings of the 21st International Symposium on Superconductivity (ISS
  2008)}\BibitemShut {NoStop}%
\bibitem [{\citenamefont {Ikeda}\ \emph {et~al.}(2016)\citenamefont {Ikeda},
  \citenamefont {Krockenberger}, \citenamefont {Irie}, \citenamefont {Naito},\
  and\ \citenamefont {Yamamoto}}]{Ikeda}%
  \BibitemOpen
  \bibfield  {author} {\bibinfo {author} {\bibfnamefont {A.}~\bibnamefont
  {Ikeda}}, \bibinfo {author} {\bibfnamefont {Y.}~\bibnamefont
  {Krockenberger}}, \bibinfo {author} {\bibfnamefont {H.}~\bibnamefont {Irie}},
  \bibinfo {author} {\bibfnamefont {M.}~\bibnamefont {Naito}}, \ and\ \bibinfo
  {author} {\bibfnamefont {H.}~\bibnamefont {Yamamoto}},\ }\href {\doibase
  10.7567/apex.9.061101} {\bibfield  {journal} {\bibinfo  {journal} {Applied
  Physics Express}\ }\textbf {\bibinfo {volume} {9}},\ \bibinfo {pages}
  {061101} (\bibinfo {year} {2016})}\BibitemShut {NoStop}%
\bibitem [{\citenamefont {Onozuka}\ \emph {et~al.}(2016)\citenamefont
  {Onozuka}, \citenamefont {Chikamatsu}, \citenamefont {Katayama},
  \citenamefont {Fukumura},\ and\ \citenamefont {Hasegawa}}]{Onozuka}%
  \BibitemOpen
  \bibfield  {author} {\bibinfo {author} {\bibfnamefont {T.}~\bibnamefont
  {Onozuka}}, \bibinfo {author} {\bibfnamefont {A.}~\bibnamefont {Chikamatsu}},
  \bibinfo {author} {\bibfnamefont {T.}~\bibnamefont {Katayama}}, \bibinfo
  {author} {\bibfnamefont {T.}~\bibnamefont {Fukumura}}, \ and\ \bibinfo
  {author} {\bibfnamefont {T.}~\bibnamefont {Hasegawa}},\ }\href {\doibase
  10.1039/C6DT01737A} {\bibfield  {journal} {\bibinfo  {journal} {Dalton
  Trans.}\ }\textbf {\bibinfo {volume} {45}},\ \bibinfo {pages} {12114}
  (\bibinfo {year} {2016})}\BibitemShut {NoStop}%
\bibitem [{\citenamefont {Anisimov}\ \emph {et~al.}(1999)\citenamefont
  {Anisimov}, \citenamefont {Bukhvalov},\ and\ \citenamefont
  {Rice}}]{Anisimov}%
  \BibitemOpen
  \bibfield  {author} {\bibinfo {author} {\bibfnamefont {V.~I.}\ \bibnamefont
  {Anisimov}}, \bibinfo {author} {\bibfnamefont {D.}~\bibnamefont {Bukhvalov}},
  \ and\ \bibinfo {author} {\bibfnamefont {T.~M.}\ \bibnamefont {Rice}},\
  }\href {\doibase 10.1103/PhysRevB.59.7901} {\bibfield  {journal} {\bibinfo
  {journal} {Phys. Rev. B}\ }\textbf {\bibinfo {volume} {59}},\ \bibinfo
  {pages} {7901} (\bibinfo {year} {1999})}\BibitemShut {NoStop}%
\bibitem [{\citenamefont {Lee}\ and\ \citenamefont {Pickett}(2004)}]{Pickett}%
  \BibitemOpen
  \bibfield  {author} {\bibinfo {author} {\bibfnamefont {K.-W.}\ \bibnamefont
  {Lee}}\ and\ \bibinfo {author} {\bibfnamefont {W.~E.}\ \bibnamefont
  {Pickett}},\ }\href {\doibase 10.1103/PhysRevB.70.165109} {\bibfield
  {journal} {\bibinfo  {journal} {Phys. Rev. B}\ }\textbf {\bibinfo {volume}
  {70}},\ \bibinfo {pages} {165109} (\bibinfo {year} {2004})}\BibitemShut
  {NoStop}%
\bibitem [{\citenamefont {Li}\ \emph {et~al.}(2019)\citenamefont {Li},
  \citenamefont {Lee}, \citenamefont {Wang}, \citenamefont {Osada},
  \citenamefont {Crossley}, \citenamefont {Lee}, \citenamefont {Cui},
  \citenamefont {Hikita},\ and\ \citenamefont {Hwang}}]{Hwang}%
  \BibitemOpen
  \bibfield  {author} {\bibinfo {author} {\bibfnamefont {D.}~\bibnamefont
  {Li}}, \bibinfo {author} {\bibfnamefont {K.}~\bibnamefont {Lee}}, \bibinfo
  {author} {\bibfnamefont {B.~Y.}\ \bibnamefont {Wang}}, \bibinfo {author}
  {\bibfnamefont {M.}~\bibnamefont {Osada}}, \bibinfo {author} {\bibfnamefont
  {S.}~\bibnamefont {Crossley}}, \bibinfo {author} {\bibfnamefont {H.~R.}\
  \bibnamefont {Lee}}, \bibinfo {author} {\bibfnamefont {Y.}~\bibnamefont
  {Cui}}, \bibinfo {author} {\bibfnamefont {Y.}~\bibnamefont {Hikita}}, \ and\
  \bibinfo {author} {\bibfnamefont {H.~Y.}\ \bibnamefont {Hwang}},\ }\href
  {\doibase 10.1038/s41586-019-1496-5} {\bibfield  {journal} {\bibinfo
  {journal} {Nature}\ }\textbf {\bibinfo {volume} {572}},\ \bibinfo {pages}
  {624} (\bibinfo {year} {2019})}\BibitemShut {NoStop}%
\bibitem [{\citenamefont {Sawatzky}(2019)}]{Sawatzky}%
  \BibitemOpen
  \bibfield  {author} {\bibinfo {author} {\bibfnamefont {G.~A.}\ \bibnamefont
  {Sawatzky}},\ }\href {https://www.nature.com/articles/d41586-019-02518-3}
  {\bibfield  {journal} {\bibinfo  {journal} {Nature}\ }\textbf {\bibinfo
  {volume} {572}},\ \bibinfo {pages} {592} (\bibinfo {year} {2019})},\ \bibinfo
  {note} {(an associate article of \cite{Hwang})}\BibitemShut {NoStop}%
\bibitem [{\citenamefont {Keimer}\ \emph {et~al.}(2015)\citenamefont {Keimer},
  \citenamefont {Kivelson}, \citenamefont {Norman}, \citenamefont {Uchida},\
  and\ \citenamefont {Zaanen}}]{KeimerReview}%
  \BibitemOpen
  \bibfield  {author} {\bibinfo {author} {\bibfnamefont {B.}~\bibnamefont
  {Keimer}}, \bibinfo {author} {\bibfnamefont {S.~A.}\ \bibnamefont
  {Kivelson}}, \bibinfo {author} {\bibfnamefont {M.~R.}\ \bibnamefont
  {Norman}}, \bibinfo {author} {\bibfnamefont {S.}~\bibnamefont {Uchida}}, \
  and\ \bibinfo {author} {\bibfnamefont {J.}~\bibnamefont {Zaanen}},\ }\href
  {https://doi.org/10.1038/nature14165 10.1038/nature14165} {\bibfield
  {journal} {\bibinfo  {journal} {Nature}\ }\textbf {\bibinfo {volume} {518}},\
  \bibinfo {pages} {179} (\bibinfo {year} {2015})}\BibitemShut {NoStop}%
\bibitem [{eca()}]{ecalj}%
  \BibitemOpen
  \href@noop {} {}\bibinfo {note} {A first-principles electronic-structure
  suite based on the PMT method, {\sf ecalj} package, is available from
  \texttt{https://github.com/tkotani/ecalj}. Its one-body part is developed
  based on the LMTO part in the LMsuit package at {\tt
  http://www.lmsuite.org/}.}\BibitemShut {Stop}%
\bibitem [{\citenamefont {Marzari}\ and\ \citenamefont
  {Vanderbilt}(1997)}]{Marzari}%
  \BibitemOpen
  \bibfield  {author} {\bibinfo {author} {\bibfnamefont {N.}~\bibnamefont
  {Marzari}}\ and\ \bibinfo {author} {\bibfnamefont {D.}~\bibnamefont
  {Vanderbilt}},\ }\href {http://prb.aps.org/abstract/PRB/v56/i20/p12847_1}
  {\bibfield  {journal} {\bibinfo  {journal} {Phys. Rev. B}\ }\textbf {\bibinfo
  {volume} {56}},\ \bibinfo {pages} {12847} (\bibinfo {year}
  {1997})}\BibitemShut {NoStop}%
\bibitem [{\citenamefont {Souza}\ \emph {et~al.}(2001)\citenamefont {Souza},
  \citenamefont {Marzari},\ and\ \citenamefont {Vanderbilt}}]{Souza}%
  \BibitemOpen
  \bibfield  {author} {\bibinfo {author} {\bibfnamefont {I.}~\bibnamefont
  {Souza}}, \bibinfo {author} {\bibfnamefont {N.}~\bibnamefont {Marzari}}, \
  and\ \bibinfo {author} {\bibfnamefont {D.}~\bibnamefont {Vanderbilt}},\
  }\href {\doibase 10.1103/PhysRevB.65.035109} {\bibfield  {journal} {\bibinfo
  {journal} {Phys. Rev. B}\ }\textbf {\bibinfo {volume} {65}},\ \bibinfo
  {pages} {035109} (\bibinfo {year} {2001})}\BibitemShut {NoStop}%
\bibitem [{\citenamefont {Aryasetiawan}\ \emph {et~al.}(2004)\citenamefont
  {Aryasetiawan}, \citenamefont {Imada}, \citenamefont {Georges}, \citenamefont
  {Kotliar}, \citenamefont {Biermann},\ and\ \citenamefont
  {Lichtenstein}}]{Aryasetiawan_2004}%
  \BibitemOpen
  \bibfield  {author} {\bibinfo {author} {\bibfnamefont {F.}~\bibnamefont
  {Aryasetiawan}}, \bibinfo {author} {\bibfnamefont {M.}~\bibnamefont {Imada}},
  \bibinfo {author} {\bibfnamefont {A.}~\bibnamefont {Georges}}, \bibinfo
  {author} {\bibfnamefont {G.}~\bibnamefont {Kotliar}}, \bibinfo {author}
  {\bibfnamefont {S.}~\bibnamefont {Biermann}}, \ and\ \bibinfo {author}
  {\bibfnamefont {A.~I.}\ \bibnamefont {Lichtenstein}},\ }\href {\doibase
  10.1103/PhysRevB.70.195104} {\bibfield  {journal} {\bibinfo  {journal} {Phys.
  Rev. B}\ }\textbf {\bibinfo {volume} {70}},\ \bibinfo {pages} {195104}
  (\bibinfo {year} {2004})}\BibitemShut {NoStop}%
\bibitem [{\citenamefont {Kotani}\ \emph {et~al.}(2007)\citenamefont {Kotani},
  \citenamefont {van Schilfgaarde},\ and\ \citenamefont
  {Faleev}}]{kotani_quasiparticle_2007}%
  \BibitemOpen
  \bibfield  {author} {\bibinfo {author} {\bibfnamefont {T.}~\bibnamefont
  {Kotani}}, \bibinfo {author} {\bibfnamefont {M.}~\bibnamefont {van
  Schilfgaarde}}, \ and\ \bibinfo {author} {\bibfnamefont {S.~V.}\ \bibnamefont
  {Faleev}},\ }\href {\doibase 10.1103/PhysRevB.76.165106} {\bibfield
  {journal} {\bibinfo  {journal} {Phys. Rev. B}\ }\textbf {\bibinfo {volume}
  {76}},\ \bibinfo {pages} {165106} (\bibinfo {year} {2007})}\BibitemShut
  {NoStop}%
\bibitem [{\citenamefont {Kotani}(2014)}]{kotani_quasiparticle_2014}%
  \BibitemOpen
  \bibfield  {author} {\bibinfo {author} {\bibfnamefont {T.}~\bibnamefont
  {Kotani}},\ }\href {\doibase 10.7566/JPSJ.83.094711} {\bibfield  {journal}
  {\bibinfo  {journal} {J. Phys. Soc. Jpn.}\ }\textbf {\bibinfo {volume}
  {83}},\ \bibinfo {pages} {094711} (\bibinfo {year} {2014})}\BibitemShut
  {NoStop}%
\bibitem [{com({\natexlab{a}})}]{comment1}%
  \BibitemOpen
  \href@noop {} {} ({\natexlab{a}}),\ \bibinfo {note} {we take $12\times 12
  \times 12(12\times 12 \times 6)$ or $10\times 10 \times 10(8\times 8 \times
  4)$ $k$-points in the LDA calculation or in the Wannierization procedure for
  LaNiO$_2$(HgBa$_2$CuO$_4$), respectively.}\BibitemShut {Stop}%
\bibitem [{\citenamefont {Bickers}\ \emph {et~al.}(1989)\citenamefont
  {Bickers}, \citenamefont {Scalapino},\ and\ \citenamefont
  {White}}]{Bickers1989}%
  \BibitemOpen
  \bibfield  {author} {\bibinfo {author} {\bibfnamefont {N.~E.}\ \bibnamefont
  {Bickers}}, \bibinfo {author} {\bibfnamefont {D.~J.}\ \bibnamefont
  {Scalapino}}, \ and\ \bibinfo {author} {\bibfnamefont {S.~R.}\ \bibnamefont
  {White}},\ }\href {\doibase 10.1103/PhysRevLett.62.961} {\bibfield  {journal}
  {\bibinfo  {journal} {Phys. Rev. Lett.}\ }\textbf {\bibinfo {volume} {62}},\
  \bibinfo {pages} {961} (\bibinfo {year} {1989})}\BibitemShut {NoStop}%
\bibitem [{\citenamefont {Bickers}\ and\ \citenamefont
  {White}(1991)}]{Bickers1991}%
  \BibitemOpen
  \bibfield  {author} {\bibinfo {author} {\bibfnamefont {N.~E.}\ \bibnamefont
  {Bickers}}\ and\ \bibinfo {author} {\bibfnamefont {S.~R.}\ \bibnamefont
  {White}},\ }\href {\doibase 10.1103/PhysRevB.43.8044} {\bibfield  {journal}
  {\bibinfo  {journal} {Phys. Rev. B}\ }\textbf {\bibinfo {volume} {43}},\
  \bibinfo {pages} {8044} (\bibinfo {year} {1991})}\BibitemShut {NoStop}%
\bibitem [{\citenamefont {Ikeda}\ \emph {et~al.}(2010)\citenamefont {Ikeda},
  \citenamefont {Arita},\ and\ \citenamefont {Kune\ifmmode~\check{s}\else
  \v{s}\fi{}}}]{Ikeda_omega0}%
  \BibitemOpen
  \bibfield  {author} {\bibinfo {author} {\bibfnamefont {H.}~\bibnamefont
  {Ikeda}}, \bibinfo {author} {\bibfnamefont {R.}~\bibnamefont {Arita}}, \ and\
  \bibinfo {author} {\bibfnamefont {J.}~\bibnamefont
  {Kune\ifmmode~\check{s}\else \v{s}\fi{}}},\ }\href {\doibase
  10.1103/PhysRevB.81.054502} {\bibfield  {journal} {\bibinfo  {journal} {Phys.
  Rev. B}\ }\textbf {\bibinfo {volume} {81}},\ \bibinfo {pages} {054502}
  (\bibinfo {year} {2010})}\BibitemShut {NoStop}%
\bibitem [{com({\natexlab{b}})}]{comment7}%
  \BibitemOpen
  \href@noop {} {} ({\natexlab{b}}),\ \bibinfo {note} {in the self-consistent
  calculation of the renormalized Green's function, the self-energy
  $\Sigma(\omega)$ at $\omega=0$ is subtracted to avoid double counting of the
  Hartree term and the static self-energy already included in the
  first-principles calculation \cite{Ikeda_omega0}. This procedure also reduces
  the tendency of the Fermi surface nesting being recovered by the
  spin-fluctuation, which in turn results in a tendency toward underestimating
  the superconducting transition temperature.}\BibitemShut {Stop}%
\bibitem [{\citenamefont {Sakakibara}\ \emph {et~al.}(2012)\citenamefont
  {Sakakibara}, \citenamefont {Usui}, \citenamefont {Kuroki}, \citenamefont
  {Arita},\ and\ \citenamefont {Aoki}}]{Sakakibara2}%
  \BibitemOpen
  \bibfield  {author} {\bibinfo {author} {\bibfnamefont {H.}~\bibnamefont
  {Sakakibara}}, \bibinfo {author} {\bibfnamefont {H.}~\bibnamefont {Usui}},
  \bibinfo {author} {\bibfnamefont {K.}~\bibnamefont {Kuroki}}, \bibinfo
  {author} {\bibfnamefont {R.}~\bibnamefont {Arita}}, \ and\ \bibinfo {author}
  {\bibfnamefont {H.}~\bibnamefont {Aoki}},\ }\href {\doibase
  10.1103/PhysRevB.85.064501} {\bibfield  {journal} {\bibinfo  {journal} {Phys.
  Rev. B}\ }\textbf {\bibinfo {volume} {85}},\ \bibinfo {pages} {064501}
  (\bibinfo {year} {2012})}\BibitemShut {NoStop}%
\bibitem [{com({\natexlab{c}})}]{comment}%
  \BibitemOpen
  \href@noop {} {} ({\natexlab{c}}),\ \bibinfo {note} {in the {\sf ecalj} code,
  VCA can be performed only for elements adjacent to each other in the periodic
  table. We have checked, with the VASP code, that the band structure is not
  strongly affected when Ba is used in place of Sr (see supplemental
  material).}\BibitemShut {Stop}%
\bibitem [{com({\natexlab{d}})}]{comment6}%
  \BibitemOpen
  \href@noop {} {} ({\natexlab{d}}),\ \bibinfo {note} {we cannot go down to
  lower temperatures in the seven- orbital model because we cannot take enough
  Matsubara frequencies required for convergence. Still, if we compare the
  results for the six- and seven-orbital models at lower temperatures taking
  the same number of Matsubara frequencies, the results are almost
  identical.}\BibitemShut {Stop}%
\bibitem [{\citenamefont {Sakakibara}\ \emph {et~al.}(2010)\citenamefont
  {Sakakibara}, \citenamefont {Usui}, \citenamefont {Kuroki}, \citenamefont
  {Arita},\ and\ \citenamefont {Aoki}}]{Sakakibara1}%
  \BibitemOpen
  \bibfield  {author} {\bibinfo {author} {\bibfnamefont {H.}~\bibnamefont
  {Sakakibara}}, \bibinfo {author} {\bibfnamefont {H.}~\bibnamefont {Usui}},
  \bibinfo {author} {\bibfnamefont {K.}~\bibnamefont {Kuroki}}, \bibinfo
  {author} {\bibfnamefont {R.}~\bibnamefont {Arita}}, \ and\ \bibinfo {author}
  {\bibfnamefont {H.}~\bibnamefont {Aoki}},\ }\href {\doibase
  10.1103/PhysRevLett.105.057003} {\bibfield  {journal} {\bibinfo  {journal}
  {Phys. Rev. Lett.}\ }\textbf {\bibinfo {volume} {105}},\ \bibinfo {pages}
  {057003} (\bibinfo {year} {2010})}\BibitemShut {NoStop}%
\bibitem [{\citenamefont {Wagner}\ \emph {et~al.}(1993)\citenamefont {Wagner},
  \citenamefont {Radaelli}, \citenamefont {Hinks}, \citenamefont {Jorgensen},
  \citenamefont {Mitchell}, \citenamefont {Dabrowski}, \citenamefont {Knapp},\
  and\ \citenamefont {Beno}}]{Wagner}%
  \BibitemOpen
  \bibfield  {author} {\bibinfo {author} {\bibfnamefont {J.}~\bibnamefont
  {Wagner}}, \bibinfo {author} {\bibfnamefont {P.}~\bibnamefont {Radaelli}},
  \bibinfo {author} {\bibfnamefont {D.}~\bibnamefont {Hinks}}, \bibinfo
  {author} {\bibfnamefont {J.}~\bibnamefont {Jorgensen}}, \bibinfo {author}
  {\bibfnamefont {J.}~\bibnamefont {Mitchell}}, \bibinfo {author}
  {\bibfnamefont {B.}~\bibnamefont {Dabrowski}}, \bibinfo {author}
  {\bibfnamefont {G.}~\bibnamefont {Knapp}}, \ and\ \bibinfo {author}
  {\bibfnamefont {M.}~\bibnamefont {Beno}},\ }\href {\doibase
  https://doi.org/10.1016/0921-4534(93)90989-4} {\bibfield  {journal} {\bibinfo
   {journal} {Physica C: Superconductivity}\ }\textbf {\bibinfo {volume}
  {210}},\ \bibinfo {pages} {447 } (\bibinfo {year} {1993})}\BibitemShut
  {NoStop}%
\bibitem [{\citenamefont {Jang}\ \emph {et~al.}(2016)\citenamefont {Jang},
  \citenamefont {Sakakibara}, \citenamefont {Kino}, \citenamefont {Kotani},
  \citenamefont {Kuroki},\ and\ \citenamefont {Han}}]{Jang_crpa}%
  \BibitemOpen
  \bibfield  {author} {\bibinfo {author} {\bibfnamefont {S.~W.}\ \bibnamefont
  {Jang}}, \bibinfo {author} {\bibfnamefont {H.}~\bibnamefont {Sakakibara}},
  \bibinfo {author} {\bibfnamefont {H.}~\bibnamefont {Kino}}, \bibinfo {author}
  {\bibfnamefont {T.}~\bibnamefont {Kotani}}, \bibinfo {author} {\bibfnamefont
  {K.}~\bibnamefont {Kuroki}}, \ and\ \bibinfo {author} {\bibfnamefont {M.~J.}\
  \bibnamefont {Han}},\ }\href {http://dx.doi.org/10.1038/srep33397} {\bibfield
   {journal} {\bibinfo  {journal} {Sci. Rep.}\ }\textbf {\bibinfo {volume}
  {6}},\ \bibinfo {pages} {33397} (\bibinfo {year} {2016})}\BibitemShut
  {NoStop}%
\bibitem [{\citenamefont {Sakakibara}\ \emph {et~al.}(2017)\citenamefont
  {Sakakibara}, \citenamefont {Jang}, \citenamefont {Kino}, \citenamefont
  {Han}, \citenamefont {Kuroki},\ and\ \citenamefont {Kotani}}]{mrpa}%
  \BibitemOpen
  \bibfield  {author} {\bibinfo {author} {\bibfnamefont {H.}~\bibnamefont
  {Sakakibara}}, \bibinfo {author} {\bibfnamefont {S.~W.}\ \bibnamefont
  {Jang}}, \bibinfo {author} {\bibfnamefont {H.}~\bibnamefont {Kino}}, \bibinfo
  {author} {\bibfnamefont {M.~J.}\ \bibnamefont {Han}}, \bibinfo {author}
  {\bibfnamefont {K.}~\bibnamefont {Kuroki}}, \ and\ \bibinfo {author}
  {\bibfnamefont {T.}~\bibnamefont {Kotani}},\ }\href {\doibase
  10.7566/JPSJ.86.044714} {\bibfield  {journal} {\bibinfo  {journal} {J. Phys.
  Soc. Jpn.}\ }\textbf {\bibinfo {volume} {86}},\ \bibinfo {pages} {044714}
  (\bibinfo {year} {2017})}\BibitemShut {NoStop}%
\bibitem [{\citenamefont {Sakakibara}\ and\ \citenamefont
  {Kotani}(2019)}]{mrpa2}%
  \BibitemOpen
  \bibfield  {author} {\bibinfo {author} {\bibfnamefont {H.}~\bibnamefont
  {Sakakibara}}\ and\ \bibinfo {author} {\bibfnamefont {T.}~\bibnamefont
  {Kotani}},\ }\href {\doibase 10.1103/PhysRevB.99.195141} {\bibfield
  {journal} {\bibinfo  {journal} {Phys. Rev. B}\ }\textbf {\bibinfo {volume}
  {99}},\ \bibinfo {pages} {195141} (\bibinfo {year} {2019})}\BibitemShut
  {NoStop}%
\bibitem [{\citenamefont {Botana}\ and\ \citenamefont {Norman}(2019)}]{Botana}%
  \BibitemOpen
  \bibfield  {author} {\bibinfo {author} {\bibfnamefont {A.~S.}\ \bibnamefont
  {Botana}}\ and\ \bibinfo {author} {\bibfnamefont {M.~R.}\ \bibnamefont
  {Norman}},\ }\href@noop {} {} (\bibinfo {year} {2019}),\ \Eprint
  {http://arxiv.org/abs/arXiv:1908.10946} {arXiv:1908.10946} \BibitemShut
  {NoStop}%
\bibitem [{\citenamefont {Shinkai}\ \emph {et~al.}(2006)\citenamefont
  {Shinkai}, \citenamefont {Ikeda},\ and\ \citenamefont {Yamada}}]{Shinkai}%
  \BibitemOpen
  \bibfield  {author} {\bibinfo {author} {\bibfnamefont {S.}~\bibnamefont
  {Shinkai}}, \bibinfo {author} {\bibfnamefont {H.}~\bibnamefont {Ikeda}}, \
  and\ \bibinfo {author} {\bibfnamefont {K.}~\bibnamefont {Yamada}},\ }\href
  {\doibase 10.1143/JPSJ.75.104712} {\bibfield  {journal} {\bibinfo  {journal}
  {Journal of the Physical Society of Japan}\ }\textbf {\bibinfo {volume}
  {75}},\ \bibinfo {pages} {104712} (\bibinfo {year} {2006})},\ \Eprint
  {http://arxiv.org/abs/https://doi.org/10.1143/JPSJ.75.104712}
  {https://doi.org/10.1143/JPSJ.75.104712} \BibitemShut {NoStop}%
\bibitem [{\citenamefont {Weber}\ \emph {et~al.}(2012)\citenamefont {Weber},
  \citenamefont {Yee}, \citenamefont {Haule},\ and\ \citenamefont
  {Kotliar}}]{Weber}%
  \BibitemOpen
  \bibfield  {author} {\bibinfo {author} {\bibfnamefont {C.}~\bibnamefont
  {Weber}}, \bibinfo {author} {\bibfnamefont {C.}~\bibnamefont {Yee}}, \bibinfo
  {author} {\bibfnamefont {K.}~\bibnamefont {Haule}}, \ and\ \bibinfo {author}
  {\bibfnamefont {G.}~\bibnamefont {Kotliar}},\ }\href
  {http://stacks.iop.org/0295-5075/100/i=3/a=37001} {\bibfield  {journal}
  {\bibinfo  {journal} {Europhys. Lett.}\ }\textbf {\bibinfo {volume} {100}},\
  \bibinfo {pages} {37001} (\bibinfo {year} {2012})}\BibitemShut {NoStop}%
\bibitem [{\citenamefont {Weber}\ \emph {et~al.}(2010)\citenamefont {Weber},
  \citenamefont {Haule},\ and\ \citenamefont {Kotliar}}]{Weber2}%
  \BibitemOpen
  \bibfield  {author} {\bibinfo {author} {\bibfnamefont {C.}~\bibnamefont
  {Weber}}, \bibinfo {author} {\bibfnamefont {K.}~\bibnamefont {Haule}}, \ and\
  \bibinfo {author} {\bibfnamefont {G.}~\bibnamefont {Kotliar}},\ }\href
  {\doibase 10.1103/PhysRevB.82.125107} {\bibfield  {journal} {\bibinfo
  {journal} {Phys. Rev. B}\ }\textbf {\bibinfo {volume} {82}},\ \bibinfo
  {pages} {125107} (\bibinfo {year} {2010})}\BibitemShut {NoStop}%
\end{thebibliography}%


\begin{thebibliography}{9}%
\makeatletter
\providecommand \@ifxundefined [1]{%
 \@ifx{#1\undefined}
}%
\providecommand \@ifnum [1]{%
 \ifnum #1\expandafter \@firstoftwo
 \else \expandafter \@secondoftwo
 \fi
}%
\providecommand \@ifx [1]{%
 \ifx #1\expandafter \@firstoftwo
 \else \expandafter \@secondoftwo
 \fi
}%
\providecommand \natexlab [1]{#1}%
\providecommand \enquote  [1]{``#1''}%
\providecommand \bibnamefont  [1]{#1}%
\providecommand \bibfnamefont [1]{#1}%
\providecommand \citenamefont [1]{#1}%
\providecommand \href@noop [0]{\@secondoftwo}%
\providecommand \href [0]{\begingroup \@sanitize@url \@href}%
\providecommand \@href[1]{\@@startlink{#1}\@@href}%
\providecommand \@@href[1]{\endgroup#1\@@endlink}%
\providecommand \@sanitize@url [0]{\catcode `\\12\catcode `\$12\catcode
  `\&12\catcode `\#12\catcode `\^12\catcode `\_12\catcode `\%12\relax}%
\providecommand \@@startlink[1]{}%
\providecommand \@@endlink[0]{}%
\providecommand \url  [0]{\begingroup\@sanitize@url \@url }%
\providecommand \@url [1]{\endgroup\@href {#1}{\urlprefix }}%
\providecommand \urlprefix  [0]{URL }%
\providecommand \Eprint [0]{\href }%
\providecommand \doibase [0]{http://dx.doi.org/}%
\providecommand \selectlanguage [0]{\@gobble}%
\providecommand \bibinfo  [0]{\@secondoftwo}%
\providecommand \bibfield  [0]{\@secondoftwo}%
\providecommand \translation [1]{[#1]}%
\providecommand \BibitemOpen [0]{}%
\providecommand \bibitemStop [0]{}%
\providecommand \bibitemNoStop [0]{.\EOS\space}%
\providecommand \EOS [0]{\spacefactor3000\relax}%
\providecommand \BibitemShut  [1]{\csname bibitem#1\endcsname}%
\let\auto@bib@innerbib\@empty
\bibitem [{eca()}]{ecalj}%
  \BibitemOpen
  \href@noop {} {}\bibinfo {note} {A first-principles electronic-structure
  suite based on the PMT method, {\sf ecalj} package, is available from
  \texttt{https://github.com/tkotani/ecalj}. Its one-body part is developed
  based on the LMTO part in the LMsuit package at {\tt
  http://www.lmsuite.org/}.}\BibitemShut {Stop}%
\bibitem [{\citenamefont {Kresse}\ and\ \citenamefont {Hafner}(1993)}]{VASP1}%
  \BibitemOpen
  \bibfield  {author} {\bibinfo {author} {\bibfnamefont {G.}~\bibnamefont
  {Kresse}}\ and\ \bibinfo {author} {\bibfnamefont {J.}~\bibnamefont
  {Hafner}},\ }\href {\doibase 10.1103/PhysRevB.47.558} {\bibfield  {journal}
  {\bibinfo  {journal} {Phys. Rev. B}\ }\textbf {\bibinfo {volume} {47}},\
  \bibinfo {pages} {558} (\bibinfo {year} {1993})}\BibitemShut {NoStop}%
\bibitem [{\citenamefont {Kresse}\ and\ \citenamefont
  {Furthm\"uller}(1996{\natexlab{a}})}]{VASP4}%
  \BibitemOpen
  \bibfield  {author} {\bibinfo {author} {\bibfnamefont {G.}~\bibnamefont
  {Kresse}}\ and\ \bibinfo {author} {\bibfnamefont {J.}~\bibnamefont
  {Furthm\"uller}},\ }\href {\doibase 10.1016/0927-0256(96)00008-0} {\bibfield
  {journal} {\bibinfo  {journal} {Comput. Mater. Sci.}\ }\textbf {\bibinfo
  {volume} {6}},\ \bibinfo {pages} {15 } (\bibinfo {year}
  {1996}{\natexlab{a}})}\BibitemShut {NoStop}%
\bibitem [{\citenamefont {Kresse}\ and\ \citenamefont {Hafner}(1994)}]{VASP2}%
  \BibitemOpen
  \bibfield  {author} {\bibinfo {author} {\bibfnamefont {G.}~\bibnamefont
  {Kresse}}\ and\ \bibinfo {author} {\bibfnamefont {J.}~\bibnamefont
  {Hafner}},\ }\href {\doibase 10.1103/PhysRevB.49.14251} {\bibfield  {journal}
  {\bibinfo  {journal} {Phys. Rev. B}\ }\textbf {\bibinfo {volume} {49}},\
  \bibinfo {pages} {14251} (\bibinfo {year} {1994})}\BibitemShut {NoStop}%
\bibitem [{\citenamefont {Kresse}\ and\ \citenamefont
  {Furthm\"uller}(1996{\natexlab{b}})}]{VASP3}%
  \BibitemOpen
  \bibfield  {author} {\bibinfo {author} {\bibfnamefont {G.}~\bibnamefont
  {Kresse}}\ and\ \bibinfo {author} {\bibfnamefont {J.}~\bibnamefont
  {Furthm\"uller}},\ }\href {\doibase 10.1103/PhysRevB.54.11169} {\bibfield
  {journal} {\bibinfo  {journal} {Phys. Rev. B}\ }\textbf {\bibinfo {volume}
  {54}},\ \bibinfo {pages} {11169} (\bibinfo {year}
  {1996}{\natexlab{b}})}\BibitemShut {NoStop}%
\bibitem [{\citenamefont {Hayward}\ and\ \citenamefont
  {Rosseinsky}(2003)}]{HaywardNd}%
  \BibitemOpen
  \bibfield  {author} {\bibinfo {author} {\bibfnamefont {M.}~\bibnamefont
  {Hayward}}\ and\ \bibinfo {author} {\bibfnamefont {M.}~\bibnamefont
  {Rosseinsky}},\ }\href {\doibase
  https://doi.org/10.1016/S1293-2558(03)00111-0} {\bibfield  {journal}
  {\bibinfo  {journal} {Solid State Sciences}\ }\textbf {\bibinfo {volume}
  {5}},\ \bibinfo {pages} {839 } (\bibinfo {year} {2003})},\ \bibinfo {note}
  {international Conference on Inorganic Materials 2002}\BibitemShut {NoStop}%
\bibitem [{\citenamefont {Jang}\ \emph {et~al.}(2016)\citenamefont {Jang},
  \citenamefont {Sakakibara}, \citenamefont {Kino}, \citenamefont {Kotani},
  \citenamefont {Kuroki},\ and\ \citenamefont {Han}}]{Jang_crpa}%
  \BibitemOpen
  \bibfield  {author} {\bibinfo {author} {\bibfnamefont {S.~W.}\ \bibnamefont
  {Jang}}, \bibinfo {author} {\bibfnamefont {H.}~\bibnamefont {Sakakibara}},
  \bibinfo {author} {\bibfnamefont {H.}~\bibnamefont {Kino}}, \bibinfo {author}
  {\bibfnamefont {T.}~\bibnamefont {Kotani}}, \bibinfo {author} {\bibfnamefont
  {K.}~\bibnamefont {Kuroki}}, \ and\ \bibinfo {author} {\bibfnamefont {M.~J.}\
  \bibnamefont {Han}},\ }\href {http://dx.doi.org/10.1038/srep33397} {\bibfield
   {journal} {\bibinfo  {journal} {Sci. Rep.}\ }\textbf {\bibinfo {volume}
  {6}},\ \bibinfo {pages} {33397} (\bibinfo {year} {2016})}\BibitemShut
  {NoStop}%
\bibitem [{\citenamefont {Sakakibara}\ and\ \citenamefont
  {Kotani}(2019)}]{mrpa2}%
  \BibitemOpen
  \bibfield  {author} {\bibinfo {author} {\bibfnamefont {H.}~\bibnamefont
  {Sakakibara}}\ and\ \bibinfo {author} {\bibfnamefont {T.}~\bibnamefont
  {Kotani}},\ }\href {\doibase 10.1103/PhysRevB.99.195141} {\bibfield
  {journal} {\bibinfo  {journal} {Phys. Rev. B}\ }\textbf {\bibinfo {volume}
  {99}},\ \bibinfo {pages} {195141} (\bibinfo {year} {2019})}\BibitemShut
  {NoStop}%
\bibitem [{\citenamefont {Wagner}\ \emph {et~al.}(1993)\citenamefont {Wagner},
  \citenamefont {Radaelli}, \citenamefont {Hinks}, \citenamefont {Jorgensen},
  \citenamefont {Mitchell}, \citenamefont {Dabrowski}, \citenamefont {Knapp},\
  and\ \citenamefont {Beno}}]{Wagner}%
  \BibitemOpen
  \bibfield  {author} {\bibinfo {author} {\bibfnamefont {J.}~\bibnamefont
  {Wagner}}, \bibinfo {author} {\bibfnamefont {P.}~\bibnamefont {Radaelli}},
  \bibinfo {author} {\bibfnamefont {D.}~\bibnamefont {Hinks}}, \bibinfo
  {author} {\bibfnamefont {J.}~\bibnamefont {Jorgensen}}, \bibinfo {author}
  {\bibfnamefont {J.}~\bibnamefont {Mitchell}}, \bibinfo {author}
  {\bibfnamefont {B.}~\bibnamefont {Dabrowski}}, \bibinfo {author}
  {\bibfnamefont {G.}~\bibnamefont {Knapp}}, \ and\ \bibinfo {author}
  {\bibfnamefont {M.}~\bibnamefont {Beno}},\ }\href {\doibase
  https://doi.org/10.1016/0921-4534(93)90989-4} {\bibfield  {journal} {\bibinfo
   {journal} {Physica C: Superconductivity}\ }\textbf {\bibinfo {volume}
  {210}},\ \bibinfo {pages} {447 } (\bibinfo {year} {1993})}\BibitemShut
  {NoStop}%
\end{thebibliography}%

\end{document}


\preprint{APS/123-QED}

\title{
Supplemental material: Model construction 
and a possibility of cuprate-like pairing in a new $d^9$ nickelate superconductor (Nd,Sr)NiO$_2$
} 

\author{Hirofumi Sakakibara}
\email{sakakibara.tottori.u@gmail.com}
\affiliation{Department of Applied Mathematics and Physics, Tottori University, Tottori 680-8552, Japan}
\affiliation{Computational Condensed Matter Physics Laboratory, RIKEN, Wako, Saitama 351-0198, Japan}
\author{Hidetomo Usui}
\affiliation{Department of Physics and Materials Science, Shimane University, 1060 Nishikawatsu-cho, Matsue, Shimane, 690-8504, Japan}
\author{Katsuhiro Suzuki}
\affiliation{Research Organization of Science and Technology, Ritsumeikan University, Kusatsu, 525-8577, Japan}
\author{Takao Kotani}
\affiliation{Department of Applied Mathematics and Physics, Tottori University, Tottori 680-8552, Japan}
\author{Hideo Aoki}
\affiliation{National Institute of Advanced Industrial Science and Technology (AIST), Tsukuba, Ibaraki 305-8568, Japan}
\affiliation{Department of Physics, The University of Tokyo, Hongo, Tokyo 113-0033, Japan}
\author{Kazuhiko Kuroki}
\affiliation{Department of Physics, Osaka University, 1-1 Machikaneyama-cho, Toyonaka, Osaka, 560-0043, Japan}

\date{\today}

\pacs{ }
\maketitle

\setcounter{equation}{0}
\setcounter{figure}{0}
\setcounter{table}{0}
\setcounter{page}{1}

\renewcommand{\theequation}{S\arabic{equation}}
\renewcommand{\thefigure}{S\arabic{figure}}
\renewcommand{\bibnumfmt}[1]{[S#1]}
\renewcommand{\citenumfont}[1]{S#1}
\renewcommand{\thetable}{S\arabic{table}}

Here we compare the band structure of the actual material (Nd,Sr)NiO$_2$, which becomes superconducting,  and those of the materials considered in the actual calculation in the main text for technical reasons.\\

\begin{center}
\textbf{LaNiO$_2$ vs. NdNiO$_2$}
\end{center}

\begin{figure}[h!]
	\includegraphics[width=7cm]{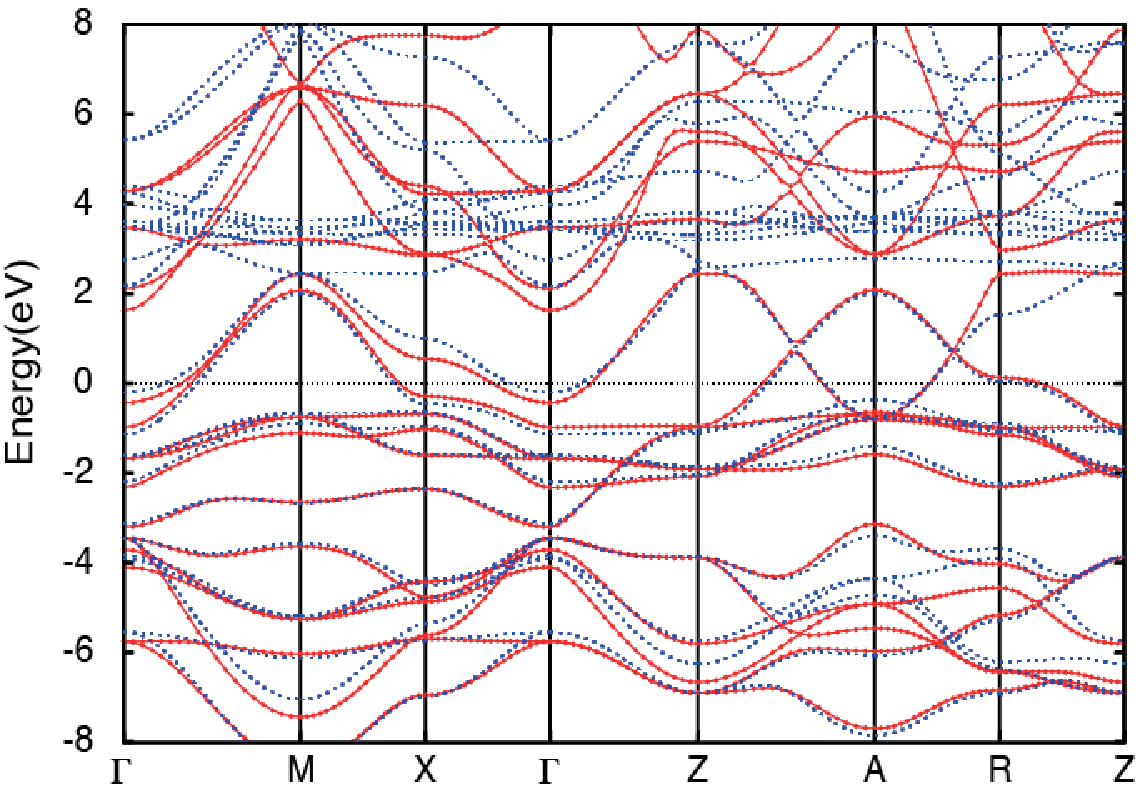}
	\caption{The first-principles band structure of LaNiO$_2$ (blue dashed lines) calculated with the {\sf ecalj} code, 
along with that of NdNiO$_2$ (red solid lines) calculated using the VASP code with the $4f$ orbitals treated as a core. In both cases, experimentally determined lattice parameters of NdNiO$_2$\cite{HaywardNd} are adopted.  
Around the Fermi energy ($E=0$) the blue and red lines 
almost overlap with each other.}
	\label{figS1}
\end{figure}

In the main text, we have considered LaNiO$_2$ instead of NdNiO$_2$ to avoid the ambiguity regarding the treatment of $4f$ orbitals. Here we show that, if we adopt the same set of lattice parameters, the band structure of LaNiO$_2$ calculated with the {\sf ecalj} code \cite{ecalj}, and that of NdNiO$_2$ calculated by the VASP code \cite{VASP1,VASP4,VASP2,VASP3}, with the $4f$ orbital treated as a core, are basically the same apart from the $f$ bands, especially around the Fermi level. Figure \ref{figS1} shows  these band structures, 
obtained for the lattice parameters of NdNiO$_2$ \cite{HaywardNd}.

\begin{center}
\textbf{(La,Sr)NiO$_2$ vs. (La,Ba)NiO$_2$}
\end{center}

\begin{figure}
	\includegraphics[width=7cm]{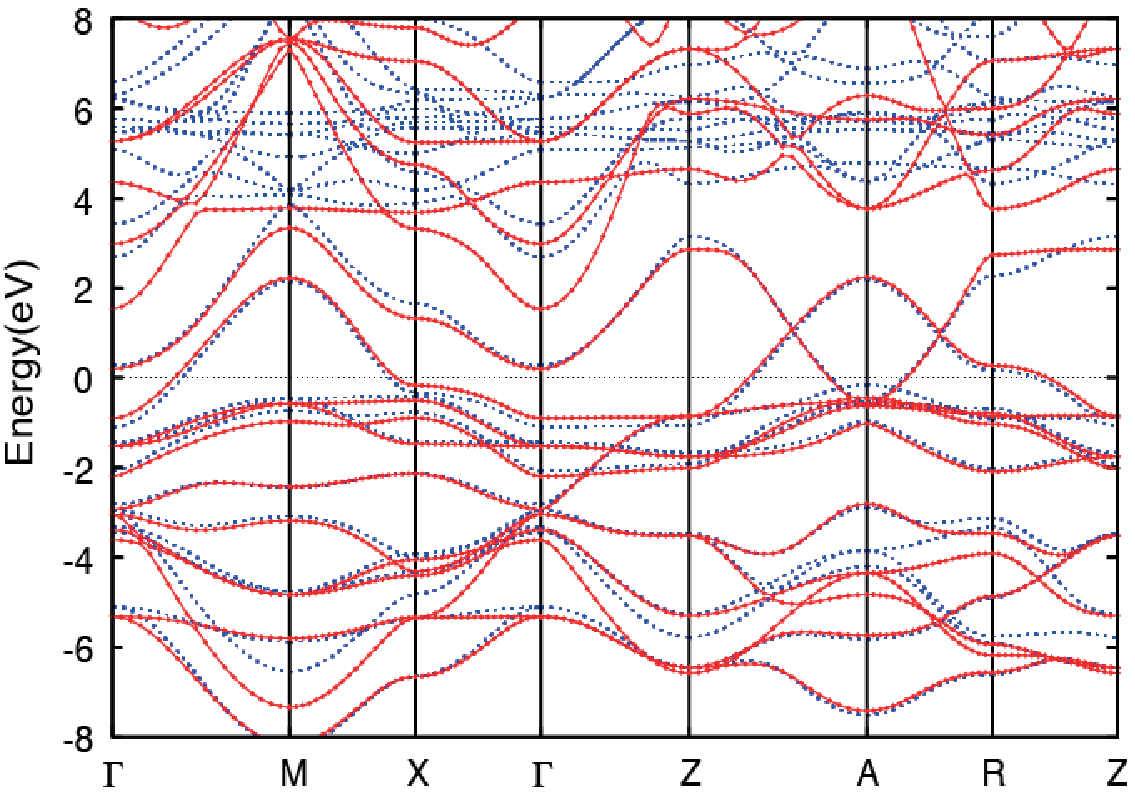}
	\caption{The first-principles band structure in 
VCA of La$_{0.8}$Ba$_{0.2}$NiO$_2$  (blue dashed lines) 
calculated with the {\sf ecalj} code \cite{ecalj}, along with that of La$_{0.8}$Sr$_{0.2}$NiO$_2$  (red solid lines)  calculated with the VASP code. In both cases, experimentally determined lattice parameters of NdNiO$_2$\cite{HaywardNd} are adopted.  
Around the Fermi energy ($E=0$) the blue and red lines 
almost overlap with each other.}
	\label{figS2}
\end{figure}

In the main text, we have treated the effect of doping within the virtual-crystal approximation, where we considered Ba instead of Sr. Here we show that the band structure calculated for (La,Ba)NiO$_2$ with the {\sf ecalj} code is essentially the same as that calculated for (La,Sr)NiO$_2$ with the VASP code. Figure \ref{figS2} shows  these band structures, obtained for the lattice parameters of LaNiO$_2$\cite{HaywardNd}.

\begin{center}
\textbf{Five-orbital model of HgBa$_2$CuO$_4$}
\end{center}

\begin{figure}
	\includegraphics[width=8cm]{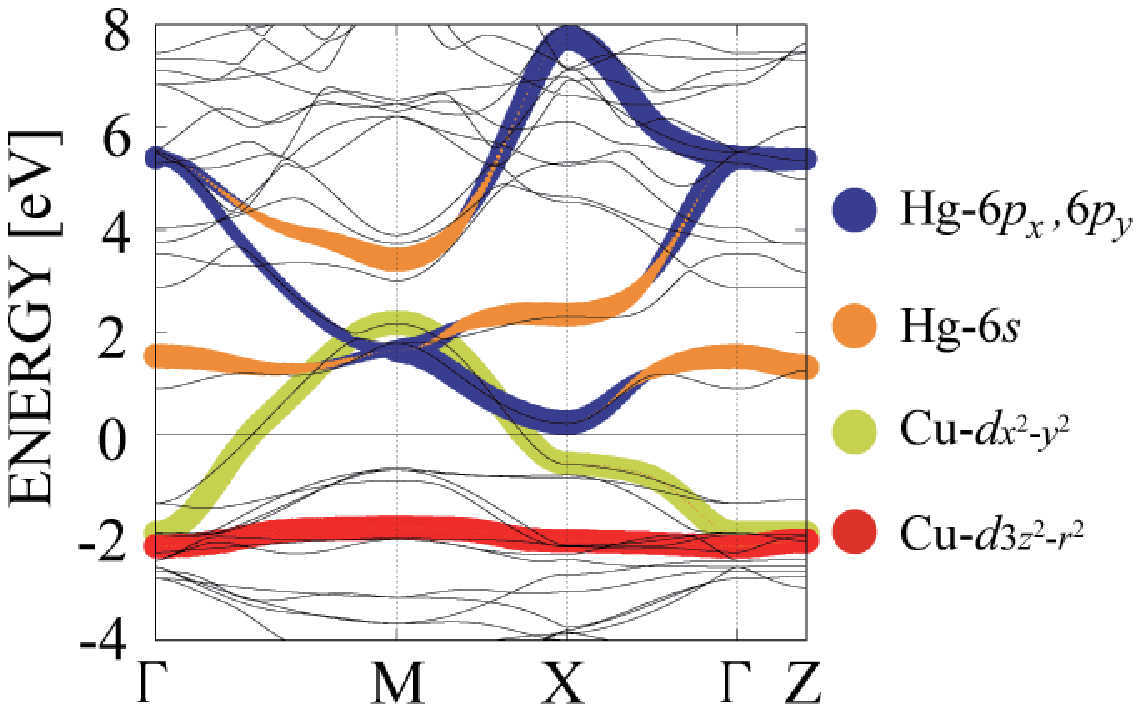}
	\caption{The first-principles band structure of HgBa$_2$CuO$_4$,
   where experimentally determined lattice parameters are adopted \cite{Wagner}.
The band structure of the five-orbital model is superposed, where 
the Wannier-orbital weight is represented by the thickness of lines with color-coded orbital characters.}
	\label{figS3}
\end{figure}

In Fig. \ref{figS3}, we present the band structure of the five orbital model of HgBa$_2$CuO$_4$, 
superposed to the first principles band structure. As seen here, although the 
Hg 6$s$-$6p_{x/y}$ bands do not intersect the Fermi level, 
they do come close to it and overlap largely with the Cu $d_{x^2-y^2}$ main band. 
This is  the reason why the explicit consideration of this orbital results in an appreciable increase of
$U_{d_{x^2-y^2}}$ from the value $U_{d_{x^2-y^2}}=2.14$ eV in Ref. \onlinecite{Jang_crpa} and \onlinecite{mrpa2}.

\begin{center}
\textbf{The Fermi surface of LaNiO$_2$ and (La,Ba)NiO$_2$}
\end{center}

\begin{figure}
	\includegraphics[width=8cm]{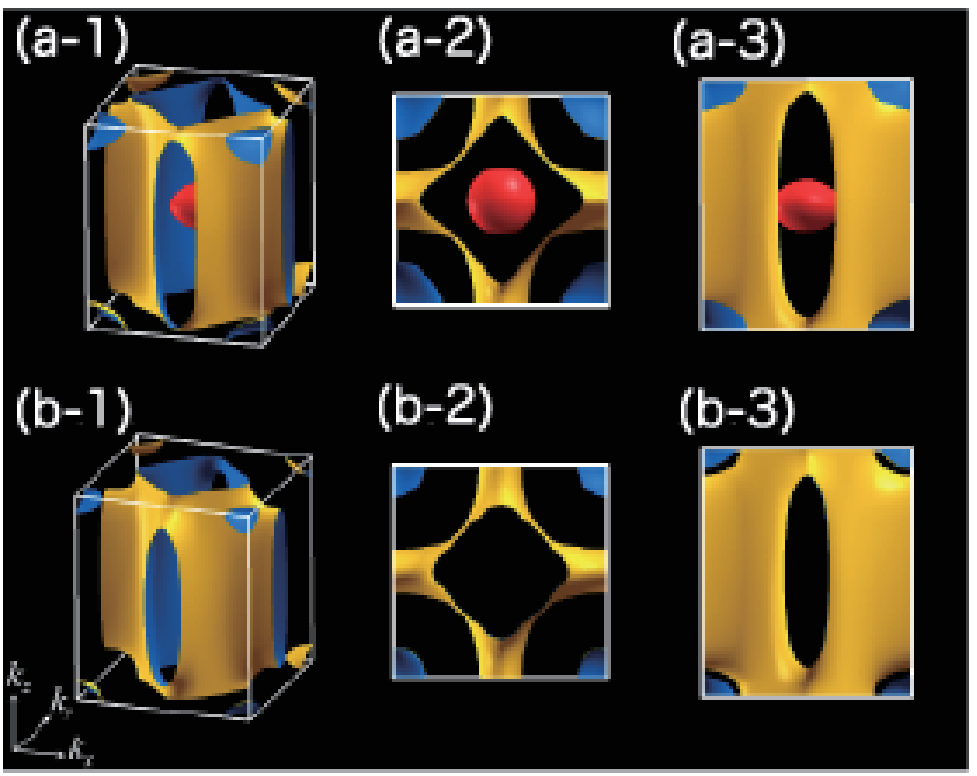}
	\caption{Three-dimensional plots of the Fermi surface. (a-1), (a-2), (a-3) are for LaNiO$_2$, while (b-1), (b-2), (b-3) are for (La,Ba)NiO$_2$. (a-2) and (b-2) are views from the $k_z$-axis direction, and 
(a-3) and (b-3) are those from the $k_x$-axis direction.}
	\label{figS4}
\end{figure}

In Fig. \ref{figS4}, we present three-dimensional plots of the Fermi surface of LaNiO$_2$ and (La,Ba)NiO$_2$. The origin of the difference in the topology of the cross sections at $k_z=0$ and $k_z=\pi$ (see the main text)  can be clearly seen here.

\begin{center}
\textbf{Interaction parameters}
\end{center}

In Table \ref{tab1}, we present all the interaction parameters evaluated for the seven-orbital model of LaNiO$_2$ and (La,Ba)NiO$_2$. Similarly in Table \ref{tab2}, those of the five-orbital model of HgBa$_2$CuO$_4$ are presented.

\begin{table}[!h]
\caption{The on-site interactions for the mother compound and $p=0.2$ doped one evaluated with cRPA. 
Here the orbital indices 1-7 indicate 
Ni $3d_{x^2-y^2}$, $3d_{3z^2-r^2}$, $3d_{xy}$, $3d_{xz}$, $3d_{yz}$, La $5d_{xy}$, and $5d_{3z^2-r^2}$ orbitals, respectively.
For example, the parameter indicated ``12''  in the Coulomb column
 is the inter-orbital interaction $U'$ given in the main text.
}
\label{tab1}
\begin{tabular}{ c  | c c |  c c} \hline
[eV] &	LaNiO$_2$	&   &    (La,Ba)NiO$_2$  & ($p=0.2$)   \\
Orbital index &	Coulomb	&	Exchange&    Coulomb  & Exchange \\\hline
11	&	3.81 	&	0.00 	&	4.19 	&	0.00 	\\
12	&	2.62 	&	0.71 	&	3.13 	&	0.73 	\\
13	&	3.30 	&	0.38 	&	3.69 	&	0.37 	\\
14	&	2.84 	&	0.66 	&	3.32 	&	0.66 	\\
15	&	2.84 	&	0.66 	&	3.32 	&	0.66 	\\
									
22	&	4.55 	&	0.00 	&	5.26 	&	0.00 	\\
23	&	2.82 	&	0.72 	&	3.34 	&	0.72 	\\
24	&	3.50 	&	0.52 	&	4.13 	&	0.53 	\\
25	&	3.50 	&	0.52 	&	4.13 	&	0.53 	\\
									
33	&	4.35 	&	0.00 	&	4.71 	&	0.00 	\\
34	&	3.06 	&	0.70 	&	3.55 	&	0.69 	\\
35	&	3.06 	&	0.70 	&	3.55 	&	0.69 	\\
									
44	&	4.56 	&	0.00 	&	5.15 	&	0.00 	\\
45	&	3.13 	&	0.70 	&	3.70 	&	0.70 	\\
									
55	&	4.56 	&	0.00 	&	5.15 	&	0.00 	\\
									
66	&	1.99 	&	0.00 	&	2.25 	&	0.00 	\\
67	&	1.52 	&	0.37 	&	1.78 	&	0.38 	\\
									
77	&	1.77 	&	0.00 	&	2.05 	&	0.00 	\\
\hline
\end{tabular}
\end{table}

 \begin{table}[!h]
\caption{The on-site interactions for HgBa$_2$CuO$_4$ evaluated with cRPA. 
Here the orbital indices 1-5 indicate 
Ni $3d_{x^2-y^2}$, $3d_{3z^2-r^2}$,  Hg $6s$,  $6p_x$, and $6p_y$ orbitals, respectively. 
}
\label{tab2}
\begin{tabular}{ c  | c c} \hline
[eV] &	 HgBa$_2$CuO$_4$  &  	\\
Orbital index &	Coulomb	& Exchange \\\hline
11	&	2.60 	&	0.00	\\
12	&	5.96 	&	0.63	\\
					
22	&	2.50 	&	0.00	\\
					
33	&	2.82 	&	0.00	\\
34	&	1.87 	&	0.22	\\
35	&	1.87 	&	0.22	\\
					
44	&	2.22 	&	0.00	\\
45	&	1.82 	&	0.41	\\
					
55	&	2.22 	&	0.00	\\
\hline
\end{tabular}
\end{table} 

\begin{figure*}
	\includegraphics[width=13cm]{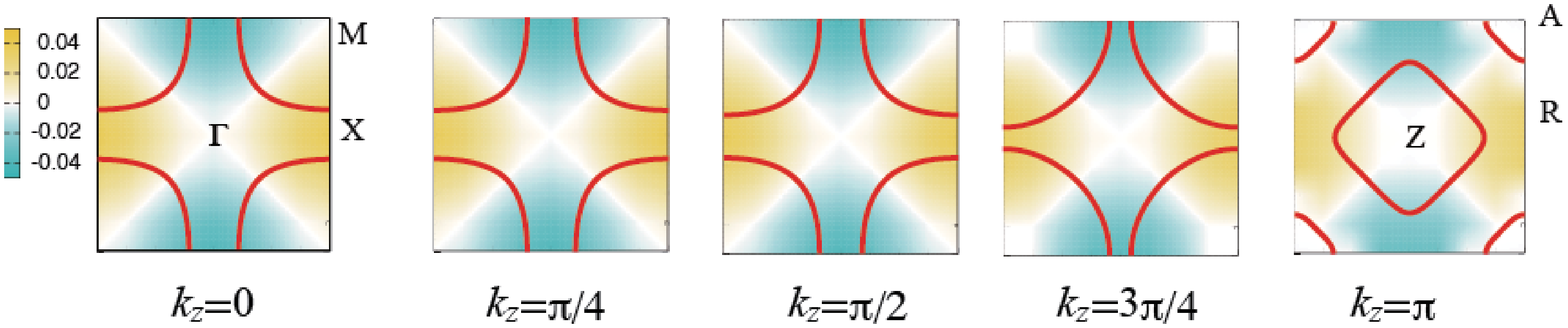}
	\caption{The FLEX gap function on band 4 (see text) of the six-orbital model of the $p=0.2$ nickelate. All the $k_z \geq 0$ cuts of the $8\times 8\times 8$ $k$-mesh are shown. $T=0.005$eV is adopted.}
	\label{figS5}
\end{figure*}

\begin{center}
\textbf{The gap function}
\end{center}
Here we present the eigenfunction of the linearized Eliashberg equation (which will be called the ``gap function'' here) for the six-orbital model of the $p=0.2$ nickelate for all the $k_z\geq 0$ cuts. When the bands are numbered in the order of the energy at each $k$ point, all portions of the Fermi surface are produced by band 4, so we plot the gap function only for this band. As seen from Fig. \ref{figS5}, the gap function on the entire Ni $3d_{x^2-y^2}$ band has $d_{x^2-y^2}$ pairing symmetry, while the gap function on the La-originating electron pocket around the A point can be barely seen. This implies that the low energy properties in the superconducting state will largely be governed by the (nearly) gapless electron pocket.

\bibliography{lanio2S}
